\documentclass{ametsocV6.1}
\usepackage{amsmath,amsfonts,amssymb,bm}
\usepackage{mathptmx}
\usepackage{newtxtext}
\usepackage{newtxmath}
\usepackage{authblk}
\usepackage{algorithm}
\usepackage{algpseudocode}
\usepackage{enumitem}
\usepackage{todonotes}
\usepackage{everypage}
\nolinenumbers

\title{Constraining Atmospheric River Uncertainty Using Instantaneous Poleward Latent Heat Transport}

\authors{Ankur Mahesh,\aff{a, b}\correspondingauthor{Ankur Mahesh, ankur.mahesh@berkeley.edu} 
William D. Collins,\aff{a, b} 
William R. Boos,\aff{a,b} 
Travis A. O'Brien,\aff{a,c} 
Yang Zhou,\aff{a} 
}

\affiliation{\aff{a}{Lawrence Berkeley National Laboratory, Berkeley, California, USA}\\
\aff{b}{University of California, Berkeley, Berkeley, California, USA}\\
\aff{c}{Indiana University, Bloomington, Bloomington, Indiana, USA}\\
}

\abstract{Atmospheric rivers (ARs) are extreme weather events that play a crucial role in the global hydrological cycle.  As a key mechanism of latent heat transport (LHT), they help maintain energy balance in the climate system. While an AR is characterized by a long, narrow corridor of water vapor associated with a low-level jet stream, there is no unambiguous definition of an AR grounded in geophysical fluid dynamics. AR identification is currently performed by a variety of threshold-based algorithms, which has introduced uncertainty in the estimated contribution of ARs to LHT. We calculate the instantaneous eddy LHT from moist, poleward anomalies.  Based on the dynamics of the large-scale atmospheric circulation, this quantity is a physics-based upper bound that constrains AR projections from the variety of detection algorithms. We quantify the contribution of ARs to transient eddies, stationary eddies, and transient-stationary eddy interactions, and we show the relative contributions of ARs vs. other processes, such as dry, equatorward transport.  We use this upper bound as a reference to quantify ARs' frequency, intensity, and temporal variability.  In the historical climate, the AR reference transport is $\sim$2.21 PW at the latitude of peak transport in Northern Hemisphere winter, with a temporal standard deviation of approximately 0.47~PW.   In a future climate projection, at this latitude, AR-induced LHT will increase by 0.5~PW and the corresponding temporal variability will increase by 0.14~PW.  The future change in the AR reference transport is larger than the changes associated with other types of atmospheric anomalies.} 

\begin{document}

\maketitle

%
%
%
%
%

%
\section{Introduction}

In the climate system, incoming solar radiation and outgoing longwave radiation are not equal at each latitude, and poleward energy transport is necessary to maintain energy balance \citep{peixoto1992physics}.  Atmospheric poleward energy transport can occur in the form of sensible heat, latent heat, and geopotential.  Atmospheric rivers (ARs) are long, narrow filaments of moisture transport that are a mechanism of poleward latent heat transport (LHT) \citep{Zhu1998}.   These extreme events can cause record-breaking floods \citep{DeFlorio2024, Corringham2019}.  For some regions, they play a crucial role as ``drought-busters,'' in which they are responsible for the vast majority of rainfall \citep{Dettinger2013}.   Despite their significance, these events do not have an unambiguous definition grounded in geophysical fluid dynamics, and various AR detection algorithms have been presented in the Atmospheric River Tracking Method Intercomparison Project (ARTMIP) \citep{Rutz2019, Shields2018}.  At a single time step, the AR detection uncertainty is visualized as a function of ARTMIP consensus in Figure~1 of \citet{Mahesh2024}.   Existing studies in ARTMIP have quantified the detection uncertainty of ARs in different reanalysis datasets \citep{Collow2022}, climate models \citep{Shields2023, OBrien2022}, and climate scenarios \citep{Shields2021, Rush2025}.  Depending on the detection algorithm used, estimates of AR-induced Integrated Vapor Transport (IVT) vary widely, ranging from 400 to 800 kg/m/s (Figure 2 of \citet{Shields2023}). This has led to significant uncertainties in the past, present, and future behavior of ARs in the midlatitudes \citep{Lora2020, Zhou2021} and polar regions \citep{Shields2022, Maclennan2025,Thaker2025}.

In this manuscript, we assess the role of ARs in poleward LHT. Traditionally, LHT is evaluated in the time mean because poleward heat transport balances incoming solar radiation and outgoing longwave radiation in a steady state system.  Time-mean LHT is expressed as contributions from transient eddies, stationary eddies, and the Mean Meridional Circulation \citep{Oort1971,holton2013introduction,Armour2019,Donohoe2020}.  Recently, \citet{Cox2024} introduced a new method to calculate poleward atmospheric heat transport instantaneously.  This enables quantification of the time variability of poleward heat transport.   Since ARs are instantaneous extreme events with a lifecycle of approximately 1 to 7 days \citep{Zhou2018}, we use this method to assess the relationship between LHT and ARs. 

Given AR detection uncertainty, we use transport from moist, poleward anomalies as the basis for a reference quantity for AR-induced transport.  The manuscript is organized into three main topics:
\begin{enumerate}
    \item AR detection uncertainty results in significant disagreement about the contribution of ARs to poleward LHT, in contrast to the original result from \citet{Zhu1998}.
    \item Eddy LHT from moist, poleward anomalies are an upper bound for AR-induced LHT across different ARTMIP detection algorithms.
    \item This upper bound is a physics-based reference quantity for ARs.  We use it to quantify ARs' temporal distribution and their future changes in frequency and intensity.
\end{enumerate}

\section{Instantaneous Poleward Latent Heat Transport}

We first summarize methods to calculate time-mean and instantaneous poleward heat transport using a dynamic perspective \citep{Armour2019}. For a full summary, we refer the reader to \citet{Cox2024}. In the time-mean, poleward LHT can be calculated as 

\begin{align}
\text{Time-Mean LHT}(\theta) &= \frac{2\pi a \cos(\theta)}{g} L_v \int_0^{P_s}  [\overline{vq}] \, dp \\
&= \frac{2\pi a\cos(\theta)}{g} L_v \int_0^{P_s}  \Bigg(
\underbrace{[\overline{v}]^\dagger[\overline{q}]^\dagger}_{\text{MMC}} 
+ \underbrace{[\bar{v}^*\bar{q}^*]}_{\text{Stationary eddy (SE)}} 
+ \underbrace{[\overline{v'^*q'^*}]}_{\text{Transient eddies (TE)}} \Bigg)  \, dp
\label{eq:standard_lht_decomposition}
\end{align}

In this equation, $\theta$ is latitude, $a$ is the radius of the Earth, $L_v$ is the latent heat of vaporization, $v$ is meridional wind, $q$ is specific humidity, $p$ is pressure, $P_s$ is surface pressure, and MMC is the mean meridional circulation. 
 Brackets $[\cdot]$ denote the zonal mean, stars $\cdot^*$ denote the deviation from the zonal mean, overbars $\overline{\cdot}$ denote the time mean, primes $\cdot'$ denote the deviation from the time mean, and swords $\cdot^\dagger$ denote the deviation from the vertical average.  In the MMC, taking the deviation from the vertical average is one possible method to account for mass conservation errors in reanalysis datasets interpolated to pressure levels \citep{Marshall2013, Cox2023, Donohoe2020}.  Handling mass conservation errors in reanalysis datasets is an important topic with significant implications for meridional heat transport \citep{Cox2023}.  Removing the vertical average in the \citet{Marshall2013} method is an alternative to the barotropic wind adjustment mass conservation method presented in \citet{Trenberth1995}. We note that the \citet{Marshall2013} method does not remove the barotropic wind component, nor any associated mass imbalance, from the TE, SE, or MT components. It is retained in the SE, TE, and MT terms that are central to the analyses here.  Some LHT decompositions use transients $[\overline{v'q'}]$ instead of transient eddies $[\overline{v'^*q'^*}]$ \citep{Trenberth1994}.  We use transient eddies, since the ``transient overturning circulation" ($\overline{[v]'[q]'}$) is orders of magnitude smaller than the other terms \citep{Donohoe2020, Marshall2013}.  Throughout this manuscript, we use the sign convention in which poleward transport is positive in the Northern Hemisphere and negative in the Southern Hemisphere.
 


Equation~\ref{eq:standard_lht_decomposition} is valid for a time-mean decomposition of poleward LHT.  Since ARs are instantaneous extreme phenomena, we calculate instantaneous poleward heat transport with the method presented in \citet{Cox2024}.

\begin{align}
\text{Instantaneous LHT}(\theta) &= \frac{2\pi a \cos(\theta)}{g} L_v \int_0^{P_s}  [vq] \, dp \\
&= \frac{2\pi a\cos(\theta)}{g} L_v \int_0^{P_s} \Bigg(\underbrace{[v]^\dagger[q]^\dagger}_{\text{MMC}} + \underbrace{[v^*q^*]}_{\text{Eddy}} \Bigg)\, dp \\
&= \frac{2\pi a\cos(\theta)}{g} L_v \int_0^{P_s}  \Bigg(
\underbrace{[v]^\dagger[q]^\dagger}_{\text{MMC}} 
+ \underbrace{[\bar{v}^*\bar{q}^*]}_{\text{SE}} 
+ \underbrace{[v'^*q'^*]}_{\text{TE}} + \underbrace{[\overline{v}^*q'^*] + [v'^*\overline{q}^*]}_{\text{Mixed Terms}} \Bigg)  \, dp
\label{eq:instantaneous_lht_decomposition}
\end{align}

The eddy transport is expanded into stationary eddies (SEs), transient eddies (TEs), and mixed terms (MTs). Compared to the time-averaged poleward heat transport, the instantaneous transport includes two new MTs that account for the interaction between stationary and transient eddies. In the time-mean, the mixed terms are zero by definition, but they can be nonzero instantaneously.  A benefit of the instantaneous perspective is that these MTs quantify the interactions between transient and stationary eddies.  Therefore, these mixed terms include the relationship between ARs and stationary features, such as orography \citep{Neiman2014}, the Aleutian Low \citep{Lora2017}, and stationary waves \citep{Menemenlis2021, Fish2022}. 

\begin{figure}[h]
\centerline{\includegraphics[width=39pc]{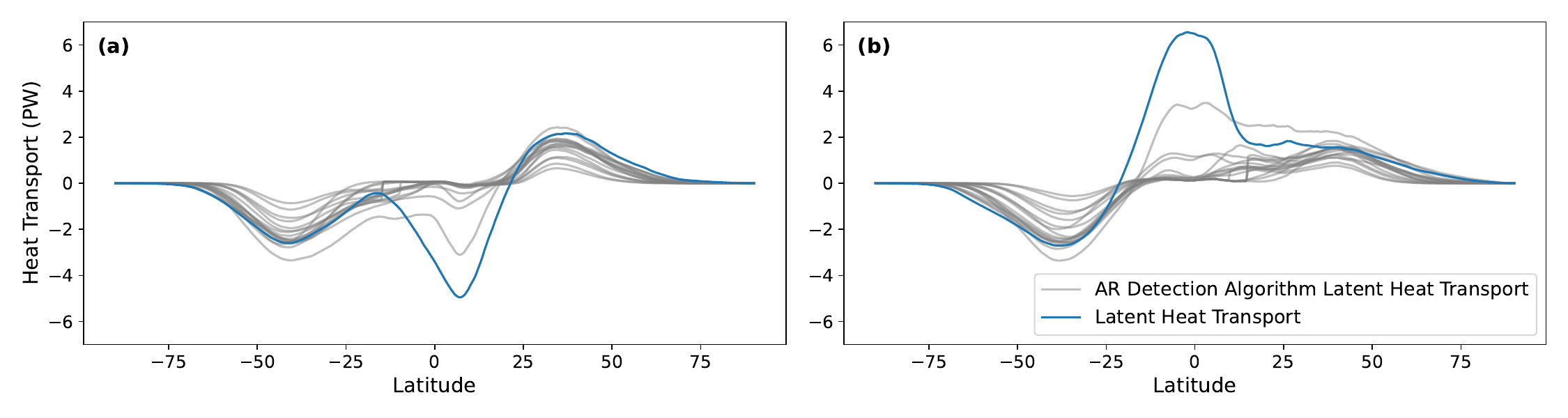}}
 \caption{\textbf{Poleward Latent Heat Transport Induced by Atmospheric Rivers (ARs)}.  Poleward LHT is shown for (a) DJF and (b) JJA from 1995-2014 MERRA2.  The poleward LHT induced by different AR detection algorithms is shown in gray.  Poleward transport is positive in the Northern Hemisphere and negative in the Southern Hemisphere, with a sign convention for heat transport being positive northward.}
 \label{fig:ar_induced_lht}
\end{figure}

\section{AR Detection Uncertainty}

In their original paper introducing a method to identify ARs from the background flow, \citet{Zhu1998} find that ARs are responsible for virtually all the extratropical poleward moisture flux.  This statement has been referred to extensively in the AR literature \citep{Gimeno2014, McClenny2020, Guan2015, Nash2018, Reid2020}, and the American Meteorological Society (AMS) Glossary of Meteorology states,  ``Horizontal water vapor transport in the midlatitudes occurs primarily in atmospheric rivers'' \citep{Ralph2018}.  However, the \citet{Zhu1998} AR detection algorithm includes uncertain thresholds, and \citet{Newman2012} show that varying the threshold parameter in the detection algorithm can lead to first-order changes in AR global coverage and moisture flux.  Since the original detection algorithm appeared in \citet{Zhu1998}, many algorithms have been introduced in ARTMIP.   These algorithms detect ARs using various input variables (such as Integrated Water Vapor (IWV) and IVT), thresholds (absolute and relative), and shape requirements.  With this comprehensive catalogue of AR detections, we revisit the statement from \citet{Zhu1998}.

 Using seventeen global detection algorithms in the ARTMIP catalog, in Figure~\ref{fig:ar_induced_lht}, we show that AR-induced LHT varies significantly by AR detection algorithm.  In this manuscript, we only use global AR detection algorithms, since global detections are necessary to assess AR contribution to zonal mean poleward LHT at all latitudes.  The algorithms used are listed in Appendix A. In December-January-February (DJF) and June-July-August (JJA), the contribution of ARs to LHT varies by almost a factor of 3.  The majority of algorithms predominantly associate ARs with extratropical LHT, although there is one outlier algorithm that also identifies significant AR activity in the tropics.   As a proportion, AR contribution to poleward LHT ranges from 15\% to approximately 105\%.\footnote{How can AR contribution to poleward LHT be greater than 100\%? Across a latitude circle in the midlatitudes, there can be equatorward heat transport \citep{Caballero2012}, which reduces the magnitude of the zonally integrated LHT.  Therefore, when algorithms only associate ARs with the poleward transport across the latitude circle, then they may find that ARs account for more than 100\% of the total heat transport.}  These results are consistent with an earlier analysis of poleward IVT by different AR algorithms (Figure 14 of \citet{Rutz2019}).  We extend the analysis in \citet{Rutz2019} to include results from more detection algorithms (seventeen analyzed here, compared to seven in \citet{Rutz2019}).  We also compare the AR-induced poleward transport to the total LHT in the climate system, to put the relative contributions of ARs in the context of this key process in the climate system.  

\section{Moist, Poleward Anomalies as an AR Reference Quantity}



The AMS Glossary AR definition identifies two characteristics that should be common across all AR definitions.  ARs should be transient phenomena, and they should be characterized by strong horizontal water vapor transport \citep{Ralph2018}.  Because of detection uncertainty with the parameters in the Zhu and Newell AR detection algorithm, \citet{Newman2012} analyze the contribution of moist ($\int_0^{P_s}q'^* dp > 0$), poleward ($\int_{700\text{hPa}} ^{P_s} v'^* dp > 0$ in the Northern Hemisphere, $\int_{700\text{hPa}} ^{P_s} v'^* dp < 0$ in the Southern Hemisphere) anomalies to moisture transport. Since ARs are associated with low-level jets and most of the moisture is concentrated in the lower troposphere, the integral of $v'^*$ is taken from 700~hPa to the surface. For ease of notation, we henceforth refer to moist, poleward anomalies as simply ($v'^* \text{poleward}, q'^* > 0$), even though we vertically integrate to identify the anomalies.

Based on the two characteristics in the AMS Glossary definition, we use moist, poleward anomalies as an upper bound on AR transport. Moist, poleward anomalies include all anomalously moist and poleward flow, so they should serve as an upper bound for phenomena characterized by strong poleward moisture transport.  This term is based on instantaneous deviations in $v$ and $q$, so it serves as an upper bound condition for transient phenomena. A limitation of this perspective is that this quantity does not include zonal moisture transport, which can also lead to high Integrated Vapor Transport, a common characteristic of ARs \citep{Shields2018}.  However, this quantity is suitable because in this manuscript, our focus is only on meridional LHT, given its well-defined equation set and implications for energy balance \citep{Armour2019}.

 


\subsection{A Method to Calculate the Transport from Moist, Poleward Anomalies}

We apply a Heaviside function, $H$, to calculate the contribution from moist, poleward anomalies to the eddy terms (TEs, SEs, and MTs in Equation~\ref{eq:instantaneous_lht_decomposition}).  We do not calculate the contribution of ARs to the MMC, as ARs are associated with eddy transport in the midlatitudes \citep{Sodemann2013, Dacre2015}.  Since ARs are long, narrow filaments with extreme moisture transport, we associate them with the eddy term in Equation~\ref{eq:instantaneous_lht_decomposition}, $[v^*q^*]$, which represents zonal mean deviations. On the other hand, the MMC represents zonal mean flow in the overturning circulation.  Additionally, in the midlatitudes, eddies are the dominant mechanism of poleward heat transport, and the MMC contribution is comparatively small \citep{Oort1971, Trenberth1994}.  Therefore, calculating AR-induced LHT solely from the eddy terms is a reasonable approximation.

To calculate the contribution of moist, poleward anomalies to TEs, we first anomalize $v$ and $q$ with respect to the time mean and zonal mean.  Then, we binarize the fields based on the existence of a moist, poleward anomaly.  For this binarization, we make use of the Heaviside function, $H$:

$
H(x) =
\begin{cases}
1, & x > 0, \\
0, & x < 0
\end{cases}
$

Finally, we calculate the AR-induced TE LHT from the moist, poleward binarized components of $v$ and $q$.  This method is summarized in Algorithm~\ref{alg:ar_te_lht}.

\newpage
\begin{algorithm}[H]
\caption{Computation of Moist, Poleward Transient Eddy Latent Heat Transport (TE LHT)}
\label{alg:ar_te_lht}
\begin{algorithmic}[1]

\Statex \textbf{Input:} \begin{itemize}
    \item $v(\theta,\phi,p,t)$ where $\theta$ is latitude, $\phi$ is longitude, $p$ is pressure, and $t$ is time
    \item $q(\theta,\phi,p,t)$
    \item  A Heaviside mask, $H(\tilde{v}'^*)$, for poleward anomalies, where
    \newline \quad \quad \quad \quad \quad $ \tilde{v}'^* = \begin{cases}\textstyle v'^*, & \theta > 0 \\ \textstyle -v'^*, & \theta < 0 \end{cases}$
 \newline \quad  \quad \quad \quad \quad $H(\tilde{v}'^*)$ is a compact notation to represent poleward anomalies ($v'^* > 0$ in Northern Hemisphere, $v'^* < 0$ in the Southern Hemisphere).
    \item Another Heaviside mask, $H(q'^*)$, for moist anomalies
\end{itemize}
\Statex \textbf{Output:} $\text{Contribution of Moist Poleward Anomalies to TE LHT }(\theta)$

\State Compute time and zonal anomalies:
\begin{align*}
v'^*(\theta,\phi,p,t) &= v(\theta,\phi,p,t) - [v](\theta,p,t) - \Big(\overline{v}(\theta,\phi,p) - \overline{[v]}(\theta,p)\Big) \\
q'^*(\theta,\phi,p,t) &= q(\theta,\phi,p,t) - [q](\theta,p,t) - \Big(\overline{q}(\theta,\phi,p) - \overline{[q]}(\theta,p)\Big)
\end{align*}

\State Split into moist, poleward components and other components using the Heaviside function:
\begin{align}
v'^*(\theta,\phi,p,t) &= v'^*(\theta,\phi,p,t) \, H(\tilde{v}'^*) H(q'^*) + v'^*(\theta,\phi,p,t)\, \big(1 - H(\tilde{v}'^*) H(q'^*)\big) \\
&= v'^*_{\text{MP}} + v'^*_{\text{Other}} \label{eq:vprime_decomp} \\
q'^*(\theta,\phi,p,t) &= q'^*(\theta,\phi,p,t) \, H(\tilde{v}'^*) H(q'^*) + q'^*(\theta,\phi,p,t)\, \big(1 - H(\tilde{v}'^*) H(q'^*)\big) \\
&= q'^*_{\text{MP}} + q'^*_{\text{Other}} \label{eq:qprime_decomp}
\end{align}

\State Compute the Moist, Poleward TE LHT. (Subscript M denotes Moist, and P denotes Poleward.)
\begin{align}
\text{Moist, Poleward TE LHT}(\theta) &= \frac{2\pi a \cos(\theta)}{g} L_v \int_0^{P_s} [v'^*_{\text{MP}} q'^*_{\text{MP}}] \, dp
\end{align}

\end{algorithmic}
\end{algorithm}

\newpage

TEs are only a function of instantaneous quantities: $v'^*$ and  $q'^*$.  However, the SEs and MTs include time-mean terms, namely $\bar{v}^*$ or $\bar{q}^*$, which require further consideration.  The time mean includes contributions from both AR and non-AR conditions.  However, our goal is to only calculate the LHT contribution from ARs (or moist, poleward anomalies, which are an AR upper bound).  To isolate the AR contribution, the Heaviside operation must be performed before the calculation of the time mean.  This order of operations is crucial because the Heaviside function is nonlinear, and it does not commute with time averaging. With this order of operations, the estimates of moist, poleward SE LHT and MT LHT are based solely on AR-like conditions, not all conditions.  We use Algorithm~\ref{alg:ar_se_lht} to calculate the AR contribution to SEs.

\begin{algorithm}[H]
\caption{Computation of Moist, Poleward Stationary Eddy Latent Heat Transport (SE LHT)}
\label{alg:ar_se_lht}
\begin{algorithmic}[1]

\Statex \textbf{Input:} \begin{itemize}[noitemsep, topsep=0pt]
    \item $v(\theta,\phi,p,t)$
    \item $q(\theta,\phi,p,t)$
    \item Heaviside mask, $H(\tilde{v}'^*)$, for poleward anomalies
    \item Heaviside mask, $H(q'^*)$, for moist anomalies
\end{itemize} 
\Statex \textbf{Output:}  Contribution of Moist, Poleward Anomalies to SE LHT

\State Compute zonal anomalies:
\begin{align*}
v^*(\theta,\phi,p,t) &= v(\theta,\phi,p,t) - [v](\theta,p,t) \\
q'^*(\theta,\phi,p,t) &= q(\theta,\phi,p,t) - [q](\theta,p,t) 
\end{align*}
\State Decompose the zonal anomalies into moist poleward terms and other components:
\begin{align*}
v^*(\theta, \phi, p, t) &=  \underbrace{v^*(\theta, \phi, p, t)  \, H(\tilde{v}'^*) \,H(q'^*) }_{\text{Moist, Poleward}} + \underbrace{v^*(\theta, \phi, p, t) \, \big (1 - H(\tilde{v}'^*) \,H(q'^*) \big ) }_{\text{Other}}   \\
q^*(\theta, \phi, p, t) &=  \underbrace{q^*(\theta, \phi, p, t)  \, H(\tilde{v}'^*) \,H(q'^*) }_{\text{Moist, Poleward}} + \underbrace{q^*(\theta, \phi, p, t) \, \big (1 - H(\tilde{v}'^*) \,H(q'^*) \big ) \ }_{\text{Other}}  
\end{align*}
\State Compute the time-mean of the zonal anomalies \textit{after} the Heaviside function has been applied.
\begin{align}
\bar{v}^*(\theta, \phi, p) &=  \overline{\underbrace{v^*(\theta, \phi, p, t)  \, H(\tilde{v}'^*) \,H(q'^*) }_{\text{Moist, Poleward}}} +\overline{\underbrace{v^*(\theta, \phi, p, t) \,\Big( 1 - H(\tilde{v}'^*) \,H(q'^*) \Big) }_{\text{Other}}}   \\
&= \bar{v}^*_{\text{MP}} + \bar{v}^*_{\text{Other}} \label{eq:vprimebar_decomp} \\
\bar{q}^*(\theta, \phi, p) &=  \overline{\underbrace{q^*(\theta, \phi, p, t)  \,H(\tilde{v}'^*) \,H(q'^*) }_{\text{Moist, Poleward}}} +  \overline{\underbrace{q^*(\theta, \phi, p, t) \,\Big( 1 - H(\tilde{v}'^*) \,H(q'^*) \Big) }_{\text{Other}}}   \\
&= \bar{q}^*_{\text{MP}} + \bar{q}^*_{ \text{Other}}  \label{eq:qprimebar_decomp}
\end{align}

\State Calculate Moist, Poleward SE LHT. (Subscript M denotes Moist, and P denotes Poleward.)
\begin{align}
\text{Moist, Poleward SE LHT}(\theta) &= \frac{2\pi a \cos(\theta)}{g} L_v \int_0^{P_s} [\bar{v}^*_{\text{MP}} \bar{q}^*_{\text{MP}}] \, dp \label{eq:mp_se_lht}
\end{align}

\end{algorithmic}
\end{algorithm}

We use analogous methods to Algorithms~\ref{alg:ar_te_lht} and \ref{alg:ar_se_lht} to calculate the moist, poleward component of the mixed terms (see Algorithm 3 in Appendix~B1).

\begin{figure}[t]
\centerline{\includegraphics[width=39pc]{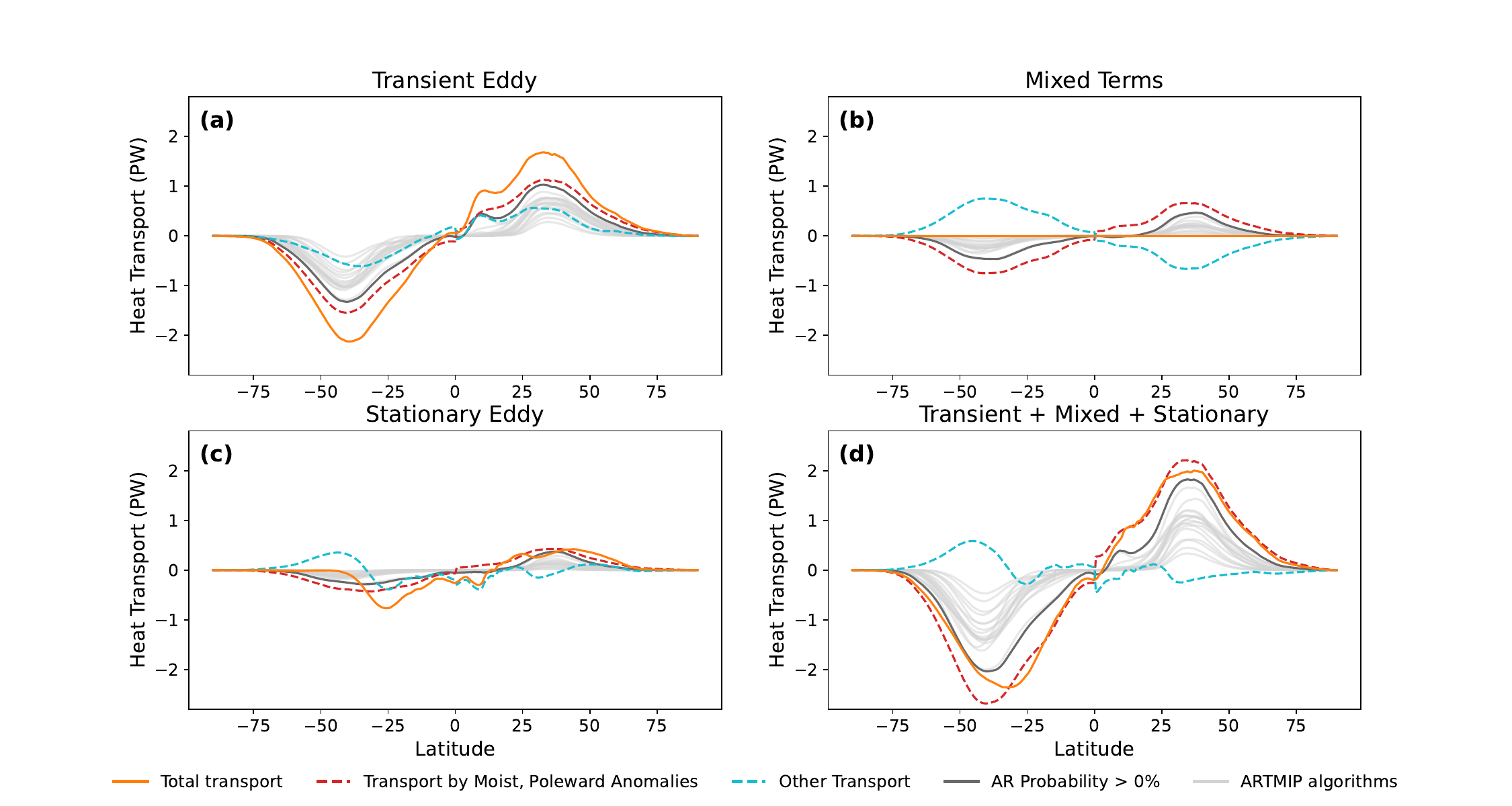}}
 \caption{\textbf{Decomposing Eddy Latent Heat Transport}. For MERRA2 DJF 1995-2014, the time-mean transient eddy heat transport, mixed terms, and stationary eddies are shown in (a), (b), and (c), respectively.  (See Equations~\ref{eq:instantaneous_lht_decomposition} and ~\ref{eq:mp_de_ds_definition} for a definition of these terms). Each of these three quantities is decomposed into an AR component (based on moist, poleward $(q'^* >= 0, v'^* >= 0)$ anomalies) and a non-AR component.  In panel (d), the sum of the transient eddies in (a), mixed terms in (b), and stationary eddies in (c) is shown to represent the total eddy contribution.  The moist, poleward transport (dotted red line) in panel (d) corresponds to the AR Reference LHT (Equation~\ref{eq:ar_reference_lht}).}
 \label{fig:djf_lht_bound}
\end{figure}

\subsection{Comparing Eddy LHT from ARTMIP Algorithms and from Moist, Poleward Anomalies}


Figure~\ref{fig:djf_lht_bound}a decomposes TE LHT into a component from moist, poleward anomalies and from all other anomalies.  These components are compared to the AR-induced TE LHT transport from nineteen ARTMIP detection algorithms.  Similar to total LHT (Figure~\ref{fig:ar_induced_lht}), there is significant detection uncertainty from ARTMIP about the role of ARs in TE LHT.  From ARTMIP, the most permissive AR detections can be derived from a condition where any algorithm identifies an AR: this condition is more permissive than any single detection algorithm.  Figure~\ref{fig:djf_lht_bound} shows this condition as ``AR Probability \textgreater\ 0\%''.  Notably, moist, poleward TE LHT is an upper bound for AR-induced LHT from all the ARTMIP algorithms. It is a tight upper bound for LHT when ``AR Probability \textgreater\ 0\%''.  These algorithms are designed to identify regions of high IVT and IWV as ARs, so most of their identifications are likely also regions of moist, poleward anomalies.

Analogous to Figure~\ref{fig:djf_lht_bound}a for TE LHT, Figure~\ref{fig:djf_lht_bound}b decomposes MT LHT into AR and non-AR transport.  As for TE LHT, there is equivalent AR detection uncertainty in the MTs, and moist, poleward anomalies serve as an upper bound for the ``AR Probability \textgreater\ 0\%'' transport.  By definition, the time-mean of the mixed terms is 0 (Figure~\ref{fig:djf_lht_bound}b). However, the time mean of the moist, poleward component is nonzero. The MTs have significant implications for ARs and LHT, as the magnitude of moist, poleward MT LHT reaches approximately 0.85~PW at the latitudinal peak in each hemisphere.  Figure~\ref{fig:djf_lht_bound}c shows the contributions of moist, poleward anomalies and AR detections from ARTMIP algorithms to SEs.  Across TEs, MTs, and SEs, AR-induced LHT has similar latitudinal structure, with a peak LHT occurring at approximately 40$^{\circ}$ in each hemisphere, increasing LHT equatorward of 40$^{\circ}$, and decreasing LHT poleward of 40$^{\circ}$.  This remains true for AR-induced SE LHT, even though the total SE LHT has a different meridional structure.  The total SE LHT has a local maximum at approximately 25$^{\circ}$ S, in the Southern Hemisphere, due to monsoonal heat transport in the summer hemisphere.  Notably, moist, poleward anomalies do not account for the local maximum of SE LHT in the subtropical high (Figure~\ref{fig:djf_lht_bound} and Appendix~Figure~A1c), indicating that other transport terms are also important for this process.  While the total LHT patterns differ between eddy types, the AR-related component exhibits a robust and consistent meridional structure. 

Of the three eddy terms, TEs have the largest magnitude of AR-induced LHT.  However, the MTs and SEs also include significant amounts of AR-induced LHT that are the same order of magnitude as the TEs.  This illustrates the importance of using instantaneous LHT from \citet{Cox2024}, since Equation~\ref{eq:standard_lht_decomposition} does not include mixed terms.  Additionally, the method in Algorithm~\ref{alg:ar_se_lht} accounts for the contribution of ARs to SEs, since ARs project onto the time-mean $\overline{v}^*$ and $\overline{q}^*$.  These approaches capture the different mechanisms through which ARs  contribute to LHT in the climate system. 

\begin{figure}[t]
\centerline{\includegraphics[width=39pc]{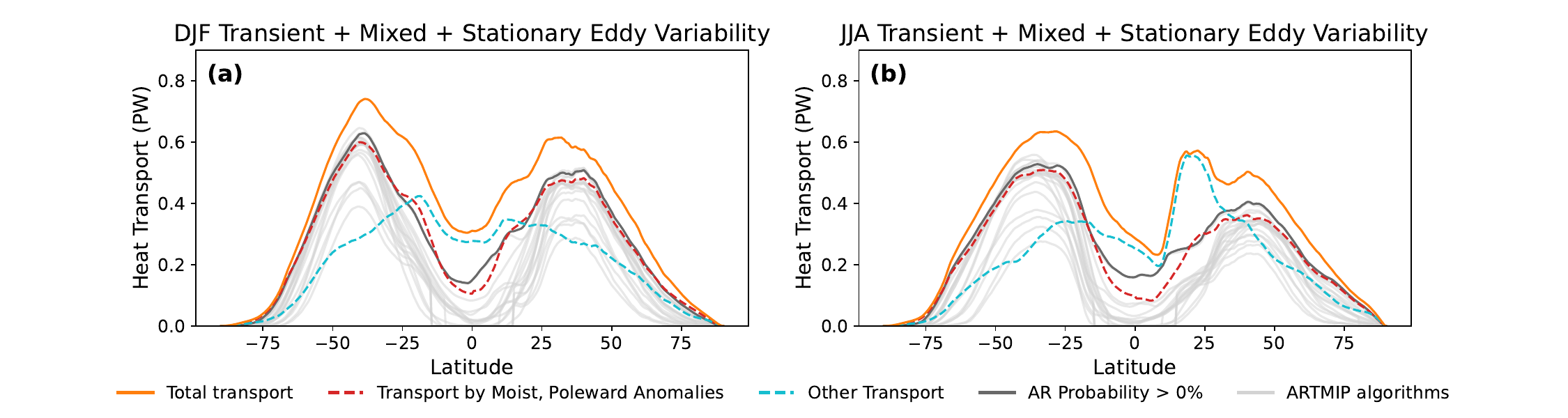}}
 \caption{\textbf{DJF Time variability of total LHT, transient eddy LHT, and AR reference LHT}.  (a) shows the DJF time variability of the sum of the transient eddies and mixed terms.  Using different detection algorithms from ARTMIP, the AR-induced quantities are shown in dark gray for each quantity.  The most permissive AR detection occurs when any detection algorithm identifies an AR.  The AR contribution from this permissive definition (``AR Probability $>$ 0\%'') is shown in dark gray. (b) is the same as (a) for JJA.  The dotted red lines (moist, poleward transport) correspond to variability in the AR Reference LHT (Equation~\ref{eq:ar_reference_lht}).}
 \label{fig:merra2_standard_deviations}
\end{figure}

A key advantage of the instantaneous LHT perspective is that it enables accurate quantification of the time variability of LHT.  While the time-mean of the MTs is zero, they have important contributions to LHT variability \citep{Cox2024}.   Figure~\ref{fig:merra2_standard_deviations} shows that the standard deviation of AR-induced LHT can reach 0.5~PW in each hemisphere.  This time variability is significant with respect to the time-mean reference AR LHT, which peaks at approximately 2.21~PW in each hemisphere (Figure~\ref{fig:djf_lht_bound}d).   This variability reflects fluctuations in zonally integrated AR-induced moisture transport across the latitude circle, and it reveals substantial time variability in AR heat transport within the climate system.  The instantaneous perspective opens up new research on what sets the mean and variability in AR-induced LHT.  

Since they are an upper bound for LHT from ARTMIP algorithms, we use moist, poleward anomalies as a physics-based reference quantity for AR-induced transport.  We define


\begin{align}
\text{AR Reference LHT}(\theta) 
&= \frac{2 \pi a \cos \theta}{g} L_v \int_0^{P_s} \bigg( 
    [v'^*_{\text{MP}}q'^*_{\text{MP}}] + [\overline{v}^*_{\text{MP}}\overline{q}^*_{\text{MP}}] \notag \\ &\quad\quad\quad\quad
    + [\overline{v}^*_{\text{MP}}q'^*_{\text{MP}}] + [v'^*_{\text{MP}}\overline{q}^*_{\text{MP}}] 
\bigg) \, dp
\label{eq:ar_reference_lht}
\end{align}

\begin{figure}[t]
\centerline{\includegraphics[width=39pc]{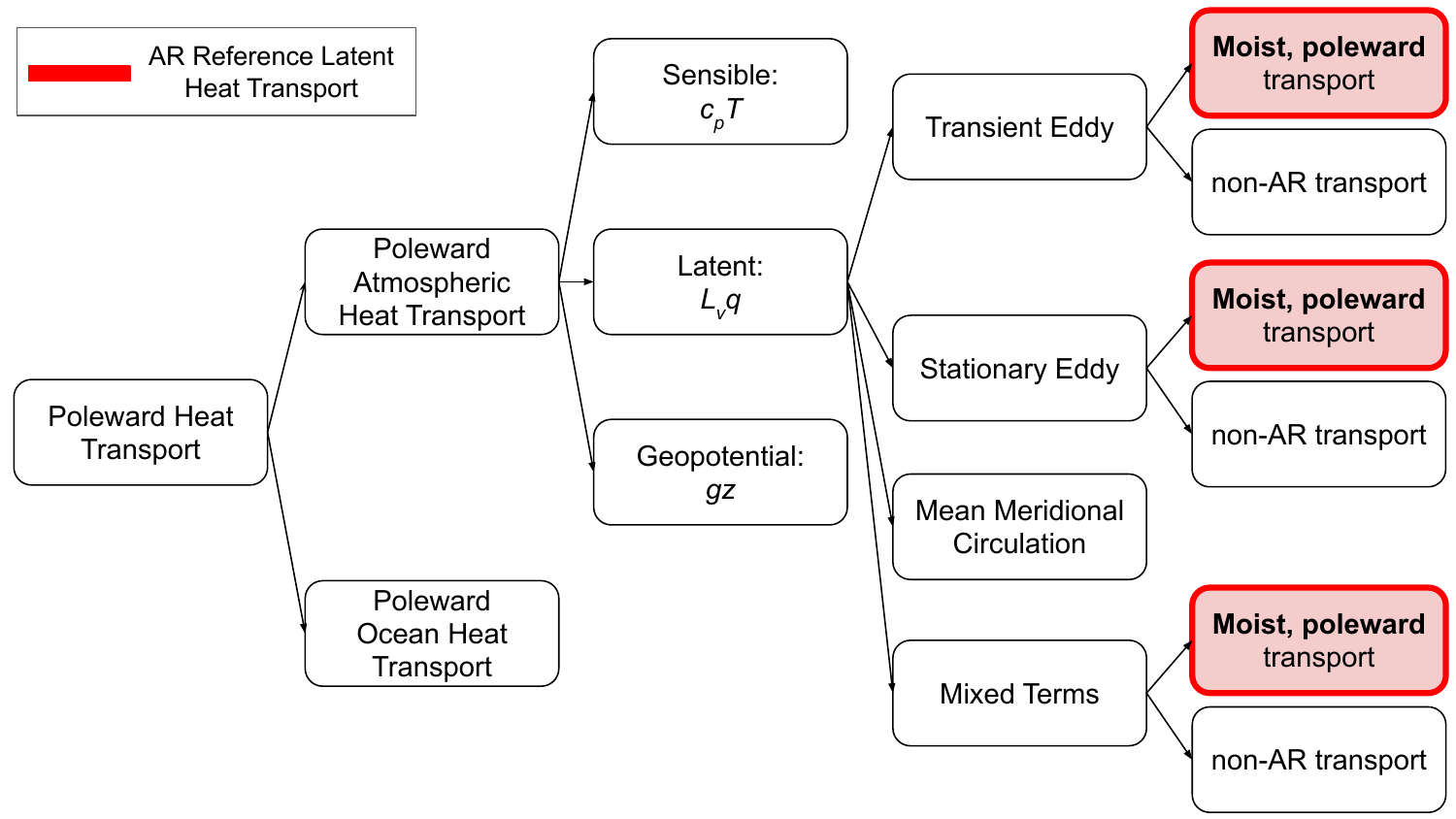}}
 \caption{\textbf{Flow Chart for Decomposing Poleward Heat Transport into Reference AR Latent Heat Transport (LHT)}.  This flowchart shows how poleward heat transport can be decomposed to derive a reference quantity for AR LHT (Equation~\ref{eq:ar_reference_lht}).  Poleward heat transport can be decomposed into atmospheric and oceanic components.  Atmospheric heat transport primarily takes the form of latent, sensible, and geopotential energy.  Poleward atmospheric LHT can be decomposed into transient eddies, stationary eddies, mixed terms, and the mean circulation (\citet{Cox2024} and Equation~\ref{eq:instantaneous_lht_decomposition} of this manuscript). Each of these terms can be decomposed into a moist, poleward component ($v'^* \text{poleward}, q'^* >0$); a dry, equatorward component ($v'^* \text{equatorward}, q'^* <0$); and a different-signs component ($v'^* \text{poleward}, q'^* <0$ or $v'^* \text{equatorward}, q'^* >0$).  The AR reference LHT is composed of moist, poleward transient eddies and mixed terms.}
 \label{fig:flow_chart}
\end{figure}

where the subscript M denotes moist, and P denotes poleward. The reference quantity for AR-induced LHT is the moist, poleward component of TEs, SEs, and MTs.  To put this quantity in the broader context of poleward heat transport, we present a flowchart in Figure~\ref{fig:flow_chart}.  This chart shows the ways to decompose poleward heat transport to derive AR reference LHT in Equation~\ref{eq:ar_reference_lht}. We show that this quantity serves as an upper bound to AR-induced transport across all ARTMIP algorithms (Figure~\ref{fig:djf_lht_bound}c).  In DJF (Figure~\ref{fig:djf_lht_bound}), the Northern Hemisphere latitude of peak AR Reference LHT is 34$^\circ$ N, which has an AR Reference LHT of 2.21~PW, with a temporal standard deviation of 0.47~PW.  The Southern Hemisphere latitude of peak AR Reference LHT is 40.5$^\circ$ S, with 2.68~PW of transport, with a temporal standard deviation of 0.59~PW.  In JJA (Appendix Figure~A1), the Northern Hemisphere latitude of peak AR Reference LHT is 43.5$^\circ$ N, which has an AR Reference LHT of 1.75~PW with a temporal standard deviation of 0.51~PW.  The Southern Hemisphere latitude of peak AR Reference LHT is 39.5$^\circ$ S, with 2.61~PW of transport, with a temporal standard deviation of 0.49~PW.

This reference quantity is unambiguously based upon the equations of heat transport (Equation~\ref{eq:instantaneous_lht_decomposition}).  Unlike AR detection algorithms, this quantity does not include uncertain IVT thresholds or shape parameters (e.g., a length scale, length-to-width ratio, or major axis length). Since this reference-based quantity does not have a shape requirement, it is not meant to identify spatially contiguous features. Instead, we use it to develop a physics-based understanding of AR frequency, intensity, and changes with climate change.  





 \subsection{Comparing Moist, Poleward TE LHT to Dry, Equatorward TE LHT}

Using the decomposition in Algorithm~\ref{alg:ar_te_lht}, TE LHT has three components: a moist, poleward ($v'^* \text{poleward}, q'^* > 0$) component; a dry, equatorward ($v'^* \text{equatorward}, q'^* < 0$) component; and a different-signs ($v'^* \text{poleward}, q'^* < 0 \cup v'^* \text{equatorward}, q'^* > 0$) component. 

\begin{align}
\text{Instantaneous TE LHT}(\theta) 
&= \frac{2 \pi a \cos \theta}{g} L_v \int_0^{P_s} [v'^*q'^*] dp \\
&= \frac{2 \pi a \cos \theta}{g} L_v \int_0^{P_s}\bigg( 
    \underbrace{[v'^*_{\text{MP}}q'^*_{\text{MP}}]}_{\text{Moist, Poleward}} 
    + \nonumber \\ &\quad\quad\quad\quad \underbrace{[v'^*_{\text{DE}}q'^*_{\text{DE}}]}_{\text{Dry, Equatorward}} + \underbrace{[v'^*_{\text{ME}}q'^*_{\text{ME}}] + [v'^*_{\text{DP}}q'^*_{\text{DP}}]}_{\text{Different-Sign transport}} 
\bigg) dp
\label{eq:mp_de_ds_definition}
\end{align}

See Appendix B2 for a full derivation. In Appendix B3, we show that SE LHT and MT LHT cannot be decomposed into the above three grouped terms. Due to the order of operations of the time averaging and Heaviside function in Section~4a, SEs and MTs are instead decomposed into nine terms.
 
For total LHT, the prevailing view is that moist, poleward anomalies \citep{Newman2012} and ARs \citep{Zhu1998,Gimeno2014,Ralph2018} play a dominant role in poleward LHT.  This can be seen in Figure~\ref{fig:djf_lht_bound}d, in which the moist, poleward transport is the dominant form of eddy transport in the midlatitudes and poles.  This underpins the conclusion that extreme events are key drivers of poleward heat transport and major precipitation events \citep{Gimeno2014, Lavers2015}.

We next compare the total eddy (TE+SE+MT) LHT (Figure~\ref{fig:djf_lht_bound}d) to the TE LHT.  Figure~\ref{fig:djf_lht_bound}a decomposes TE LHT into a Moist, Poleward component and Other component.   For TEs, the Other component is the sum of Dry, Equatorward transport and Different-Signed transport (Equation~\ref{eq:mp_de_ds_definition}).

\begin{figure}[t]
\centerline{\includegraphics[width=39pc]{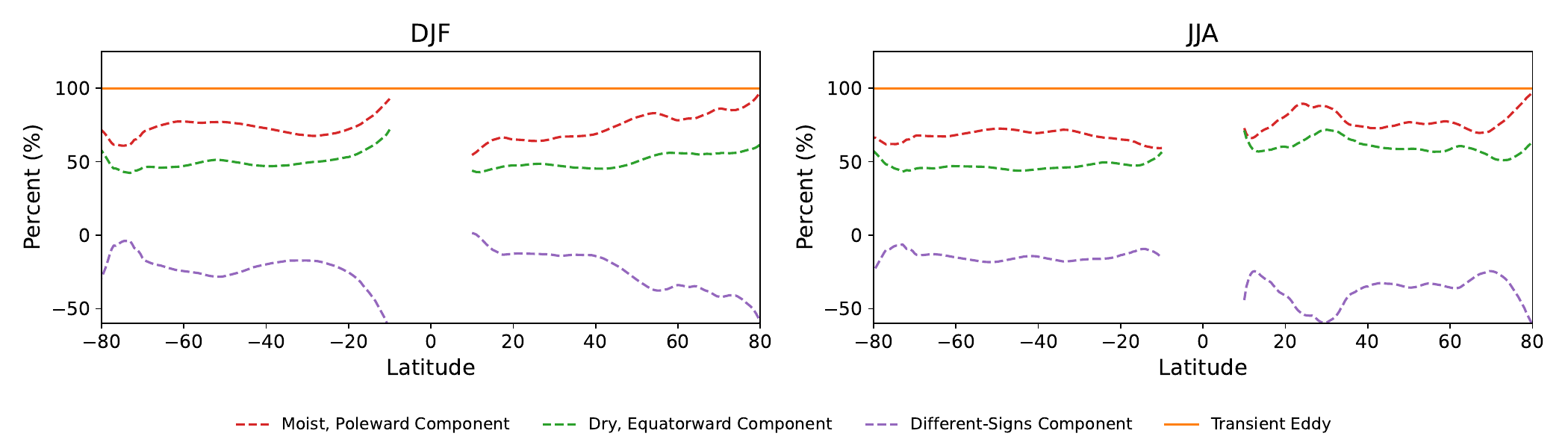}}
 \caption{\textbf{Percent Contributions of Moist Poleward, Dry Equatorward, and Different Sign $v$ and $q$ anomalies to transient eddy transports.}  In MERRA2 (1995-2014), the proportional contributions of moist, poleward; dry, equatorward; and different-sign anomalies are shown to transient eddy LHT.  Contributions are shown for DJF (left) and JJA (right). The absolute contributions are shown in Appendix~Figure~A2.}
\label{fig:relative_contributions_mp_de}
\end{figure}

Unlike total eddy LHT, moist, poleward anomalies do not dominate TE LHT. Figure~\ref{fig:relative_contributions_mp_de} shows that moist, poleward and dry, equatorward anomalies have the same order of magnitude in TE LHT. While moist, poleward anomalies do have larger LHT, the dry, equatorward transport has the same order of magnitude and is not negligible in comparison. In the extratropics, the contributions of moist, poleward anomalies and dry, equatorward anomalies to TE LHT are comparable; their relative contributions to TE LHT are within 25\%. Appendix~Figure~A2 shows the absolute magnitudes. Moist, poleward anomalies are an upper bound for AR-related transport, but dry, equatorward anomalies are not indicative of AR-related transport. Therefore, other, non-AR phenomena play an important role in poleward TE LHT.  

Why do moist, poleward anomalies dominate eddy LHT, but not TE LHT? For TEs, moist, poleward anomalies and dry, equatorward anomalies have the same sign, and they both result in poleward transport with equal magnitudes.   In contrast, for MTs, moist, poleward anomalies have the opposite sign as Other transport (Figure~\ref{fig:djf_lht_bound}b). Therefore, moist, poleward MTs result in net poleward transport, and Other MTs result in equatorward transport.  When combining TEs, SEs, and MTs, the equatorward transport from Other MTs offsets the poleward transport from Other TEs. Consequently, MTs play a key role in determining the net total AR-induced LHT. While moist, poleward anomalies are dominant in total LHT, their contribution to TE LHT is reduced due to the compensating effects of the MTs.


\section{Symmetric and Long-Tailed Distributions of AR Latent Heat Transport}

Traditionally, most of the poleward LHT is thought to occur within extreme events, in which a few events carry the bulk of the LHT for the season \citep{Zhu1998, Messori2012}.  In the zonal mean across a latitude circle, however, \citet{Cox2024} show that the temporal distribution of poleward moist static energy transport is not heavily skewed.  This indicates that a focus on mean and variance is more important than one on extreme events.  They analyze poleward moist static energy transport, which includes the sum of geopotential, latent heat, and sensible heat.  They discuss that extreme events may be more significant at the grid cell level, but in the zonal mean, moist static energy transport has a nearly symmetric (not long-tailed) temporal distribution.  Using this distinction between grid-cell and zonal mean transport, we reconcile the oft-cited dominant role of ARs in LHT with the \citet{Cox2024} result. 

\begin{figure}[t]
\centerline{\includegraphics[width=39pc]{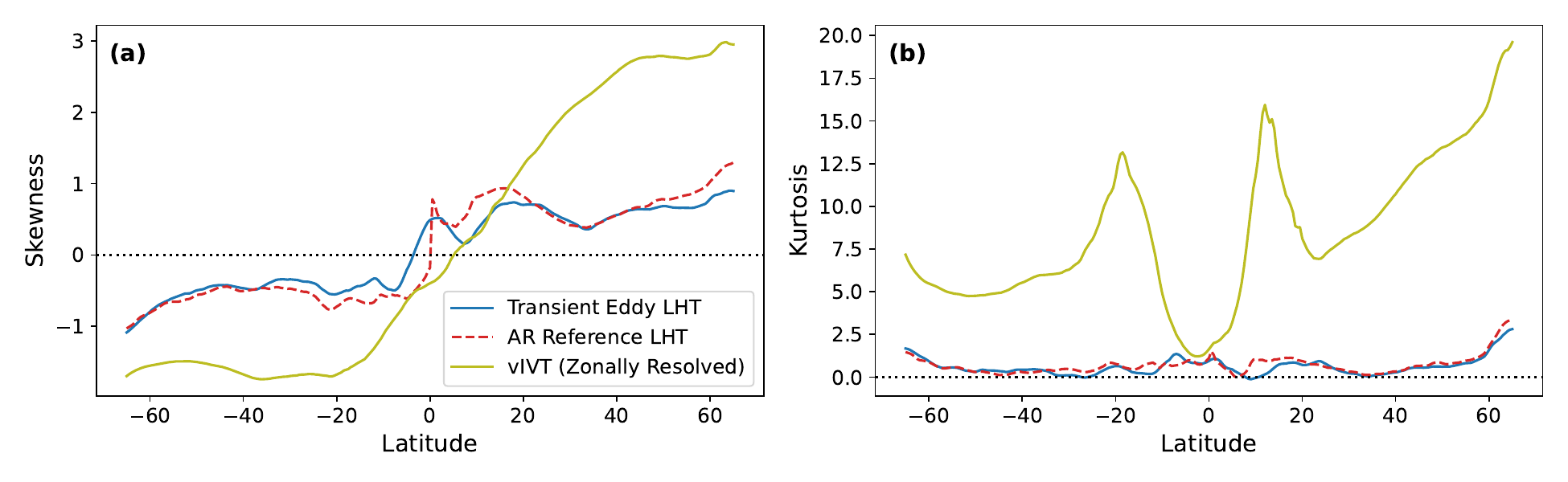}}
 \caption{\textbf{DJF Temporal Distribution of total LHT, transient eddy LHT, AR upper bound LHT, and northward IVT}.  At each latitude, the skewness (a) and kurtosis (b) of the temporal distribution are shown for total LHT, transient eddy LHT, AR upper bound LHT, and northward IVT.  The former three quantities are zonal mean quantities, whereas the northward IVT is a zonally resolved quantity.  The AR upper bound LHT is the sum of transient eddies and mixed terms, given a moist, poleward anomaly. The temporal distribution is calculated across DJF 1995-2014 in MERRA2.  Skewness and kurtosis are 0 for a Gaussian distribution.}
 \label{fig:djf_skewness_kurtosis}
\end{figure}

At each latitude, we calculate the distribution of meridional IVT (vIVT, where v refers to meridional wind) across all grid cells and times.  We note that vIVT is a zonally resolved quantity, not a zonal mean across a latitude circle.   Figure~\ref{fig:djf_skewness_kurtosis} shows that the DJF distribution of the vIVT is heavily skewed and non-Gaussian, with skewness and kurtosis values that are significantly greater than 0.  The same is true of skewness and kurtosis in JJA (Appendix~Figure~A3).  However, the distributions of TE LHT and AR reference LHT have lower skewness and kurtosis values. At latitudes of 40, 50, and 60 degrees North in DJF, we show the shape of these distributions in Appendix~Figure~A4.  

At the grid cell level, extreme events, such as ARs, account for the bulk of the transport: vIVT is heavily skewed and has a high kurtosis.  However, there are typically 4–8 ARs occurring simultaneously in each hemisphere (Figure 6 of \citet{Ullrich2021}, Figure~8 of \citet{OBrien2020Detection}, Figure~3 of \citet{Mahesh2024}, Section 1 and Figure~2 of \citet{Zhu1998}).  Across the latitude circle, the AR transport yields a more symmetric temporal distribution of LHT (less skew) and is less heavy-tailed (less kurtosis). The key distinction is that the zonal mean transport is less skewed and leptokurtic than the zonally resolved vIVT.  However, the former still contains nonzero values of these quantities and is not entirely Gaussian. The skewness and kurtosis reported here for TE LHT are similar to the corresponding values for TE moist static energy transport in \citet{Cox2024}


\section{Future Changes in ARs}

\subsection{Latent Heat Transport}


As in reanalysis, there is significant AR detection uncertainty in future projections of ARs and their LHT.  In this section, we assess the relationship between ARs and LHT in the MRI-ESM2.0 earth system model \citep{Yukimoto2019}, and we use the AR reference LHT as a benchmark estimate of future AR changes. Catalogs from six AR detection algorithms are available for the MRI model as part of the ARTMIP Tier 2 experiment \citep{OBrien2022}.   The uncertainty from these six algorithms is similar to the uncertainty from the nineteen catalogs available in reanalysis (Appendix~Figure~A5 and Figure~\ref{fig:djf_lht_bound}c).  Across scenarios (historical and SSP5-8.5 \citep{ONeill2016}) and seasons (DJF and JJA), moist, poleward anomalies serve as an upper bound for AR-induced TE LHT and MT LHT in the MRI simulations (Appendix~Figure~A5).  Since this model simulates both historical and future climate, we use these anomalies as a physics-based reference quantity to study future AR changes.

Due to Clausius-Clapeyron scaling, the tropical moisture increase is greater than the polar increase, resulting in a steeper equator-to-pole moisture gradient under future climate change (Fig. S4 of \citet{Hahn2021}).  This tropical amplification of moisture contrasts with the polar amplification of near-surface air temperature \citep{Holland2003, Smith2019}.  Poleward heat transport has been approximated as a downgradient diffusive process \citep{Armour2019, Hwang2010, Siler2018}.  With a steeper equator-to-pole moisture gradient, the diffusive perspective implies an increase in poleward LHT.  We explore the magnitude and latitudinal variation of this increase in the context of ARs. 

\begin{figure}[t]
\centerline{\includegraphics[width=39pc]{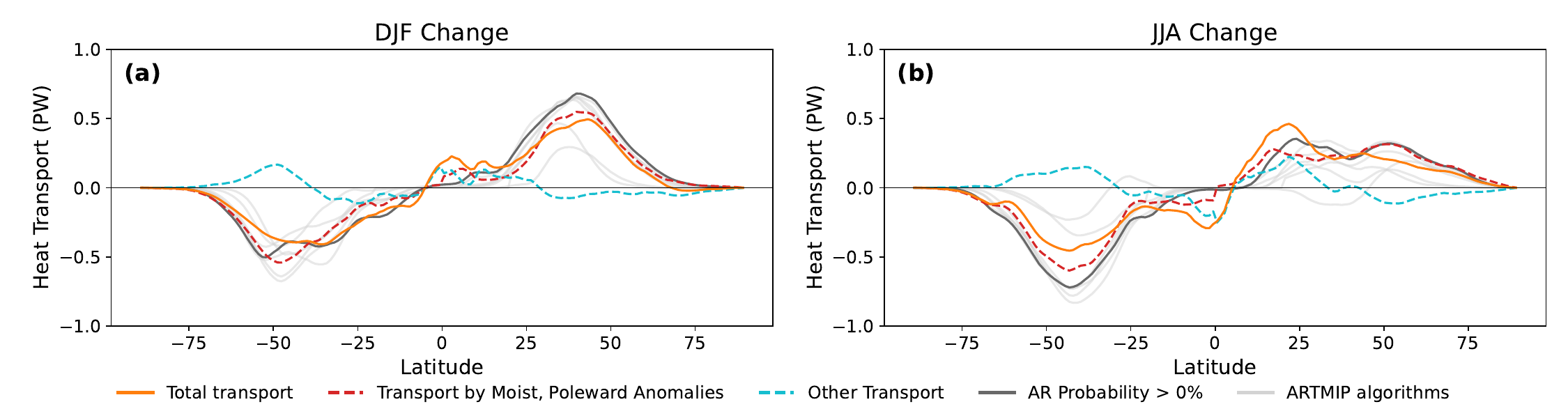}}
 \caption{\textbf{Climate Change in AR-Induced Latent Heat Transport}.  The effect of climate change on AR-induced LHT is calculated as the difference between  2081-2100 in the SSP5-8.5 scenario and 1995-2014 in the historical scenario simulated by the MRI-ESM2-0  model. The change in AR-induced latent heat transport is shown for transient eddy, stationary eddy, and mixed term (TE+SE+MT) latent heat transport (LHT). The change in TE+SE LHT is decomposed into the changes from transport by moist, poleward anomalies and transport by other anomalies.   The changes in TE+SE+MT LHT from individual ARTMIP detection algorithms are shown in light gray. Results are shown for DJF (a) and JJA (b).  The moist, poleward transport (dotted red line) corresponds to the time variability in AR Reference LHT (Equation~\ref{eq:ar_reference_lht}).}
 \label{fig:ar_intensity_change}
\end{figure}

Figure~\ref{fig:ar_intensity_change} shows the DJF and JJA changes in the AR reference LHT, comparing the 2081-2100 period under the SSP5-8.5 scenario to the 1996-2014 historical scenario. In DJF, the AR Reference LHT increases by 0.5~PW at 34$^\circ$ N and by 0.43~PW at 40.5$^\circ$ S.  In JJA, the AR Reference LHT increases by 0.25~PW at 43.5$^\circ$ N and 0.56~PW at 39.5$^\circ$ S.  These latitudes are the latitudes of peak transport in the Northern and Southern Hemispheres discussed in Section~3a. By definition, the historical average of the MTs is 0, and the climate change from the MTs is 0.  (The moist, poleward component of the historical and future MTs can be nonzero, however.)  Therefore, the future climate change in the total component comes from TEs and SEs.  In the Northern Hemisphere summer, the increase in TE and AR reference LHT is muted due to the increase in stationary eddy transport \citep{Donohoe2020}.  Across all latitudes and both DJF and JJA, the change in moist, poleward transport is greater than the change in other transport (Figure~\ref{fig:ar_intensity_change}).  This indicates that the AR Reference LHT is an important driver of future changes in eddy LHT.

Figure~\ref{fig:ar_intensity_change} shows that the ARTMIP algorithms have a large spread in AR-induced LHT under future climate change.  The AR reference LHT provides an unambiguous framework to assess future changes in AR-induced LHT (Figure~\ref{fig:ar_intensity_change}).  The future change in AR LHT from some ARTMIP algorithms can exceed the change in future AR Reference LHT. This suggests that, under warming, AR LHT more closely approaches the AR reference LHT upper bound.  In the future, a greater fraction of the reference LHT is being classified as AR transport by these algorithms.   This behavior is consistent with the thermodynamic expectation from Clausius-Clapeyron scaling: as temperatures rise, the atmosphere can hold more moisture \citep{Held2006}. The increased moisture can also amplify the intensity and spatial extent of moisture fluxes, making it more likely that they exceed the thresholds used by ARTMIP algorithms; this can increase the amount of transport identified as AR \citep{OBrien2022,Lavers2013,Shields2023}.

\begin{figure}[H]
\centerline{\includegraphics[width=39pc]{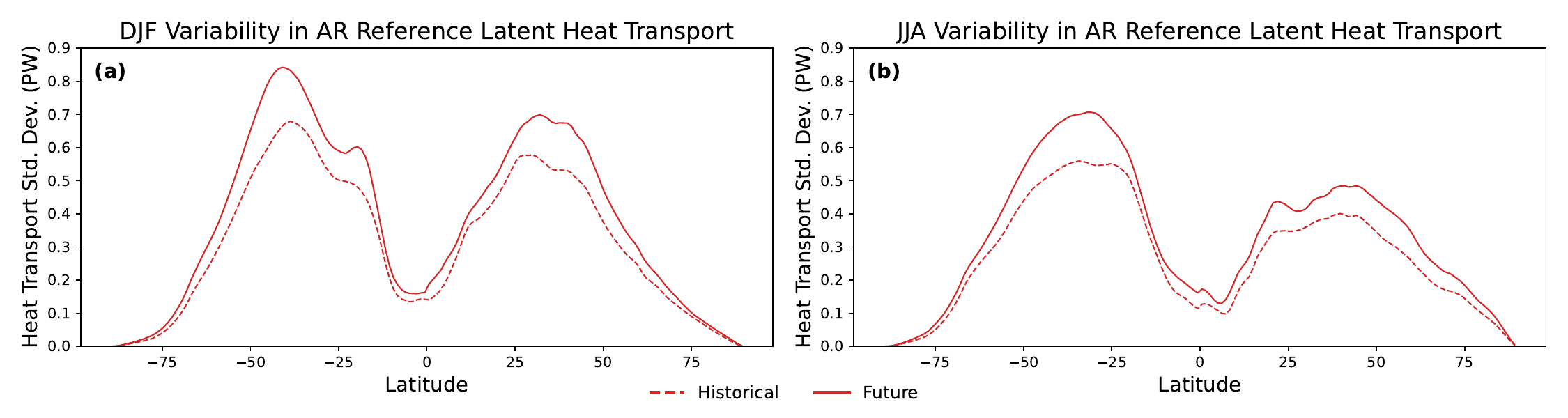}}
 \caption{\textbf{Climate Change in Variability of AR Reference Latent Heat Transport (LHT)}. AR reference LHT is the moist, poleward component of transient eddy (TE), stationary eddy (SE), and mixed term (MT) LHT (Equation~\ref{eq:ar_reference_lht}). In the MRI earth system model, the historical (1995-2014) temporal variability of AR reference LHT is shown in the dashed red line.  It is the MRI analogue to the moist, poleward component of TE, SE, and MT LHT transport in MERRA2 reanalysis (Figure~\ref{fig:djf_lht_bound}d and Figure~A1d).  The MRI future (2081-2100) SSP5-8.5 simulation is shown in the solid red line for DJF (a) and JJA (b).  AR Reference LHT is defined in Equation~\ref{eq:ar_reference_lht}. }
 \label{fig:ar_variability_historical_future}
\end{figure}

The historical variability of AR reference LHT is similar between MERRA2 (Figure~\ref{fig:merra2_standard_deviations}) and the MRI earth system model (Figure~\ref{fig:ar_variability_historical_future}).  Under future climate conditions, the temporal variability of AR reference LHT increases (Figure~\ref{fig:ar_variability_historical_future}).  This increase is approximately 0.1-0.2~PW in the extratropics during both seasons. We discuss possible relations between the increase in variability and the meridional moisture gradient in the Discussion.

\newpage
\begin{figure}[H]
\centerline{\includegraphics[width=21pc]{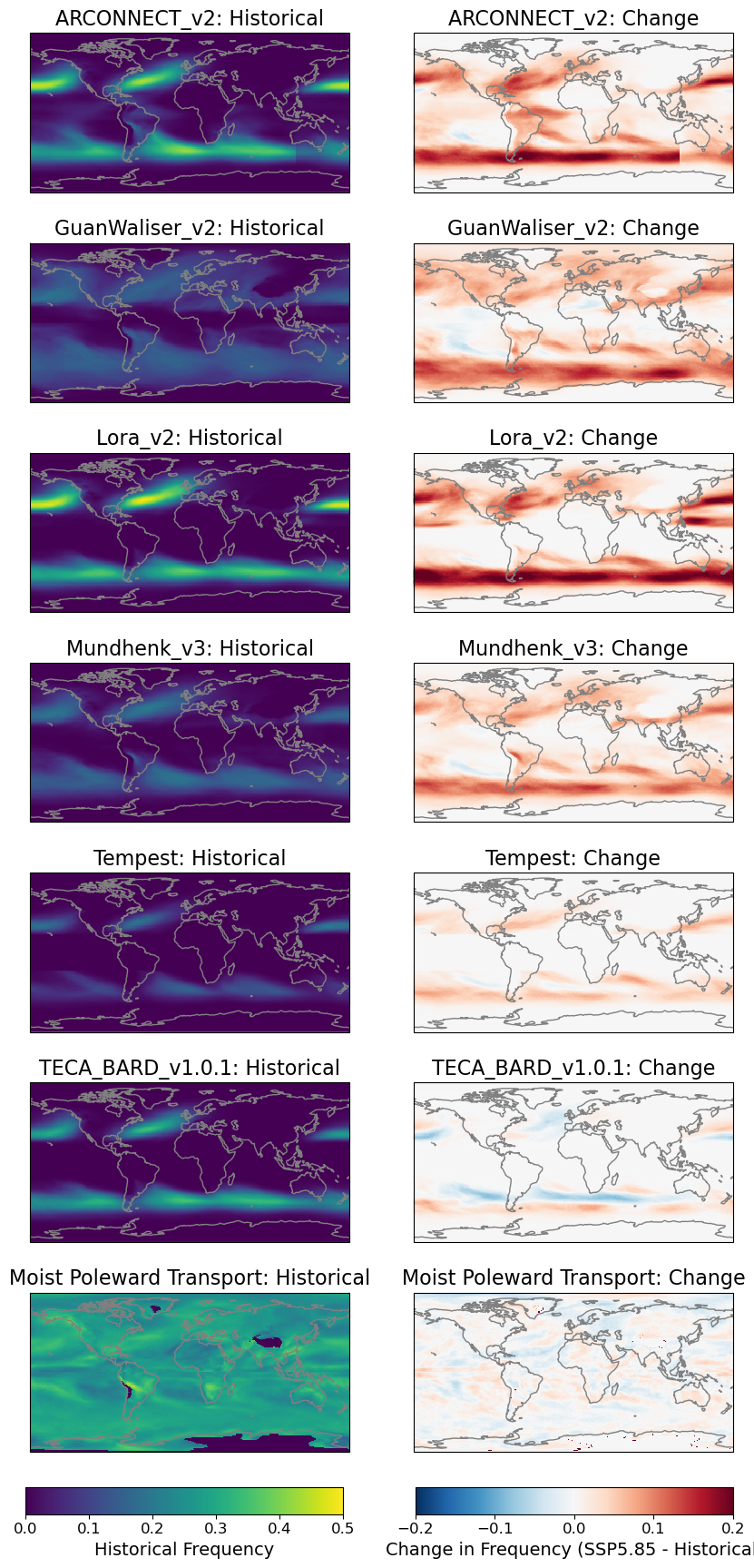}}
 \caption{\textbf{DJF Climate Changes in Frequency of ARs and Moist, Poleward Anomalies}. The left column shows the AR frequency for six different global ARTMIP algorithms (ARCONNECT, Guan-Waliser, Lora, Mundhenk, Tempest, and TECA\_BARD).  The right column shows the change in frequency calculated between the DJF 2081-2100 in the SSP5-8.5 and the DJF 1995-2014 historical period simulated by the MRI-ESM2-0 climate model. The bottom row shows the historical frequency and future change in moist, poleward $(v'^*>0 \text{ \& }q'^*>0)$ anomalies.}
 \label{fig:djf_ar_frequency}
\end{figure}

\subsection{Frequency}

Moist, poleward anomalies also serve as a reference estimate for historical and future changes in AR frequency.  Figure~\ref{fig:djf_ar_frequency} and Appendix~Figure~A6 show that five out of the six algorithms identify robust increases in AR frequency under future climate.  At most latitudes in the extratropics, the AR frequency increase nears or surpasses 10\%. In DJF,  the remaining algorithm (TECA\_BARD\_v0.1.0) identifies a poleward shift of ARs in the Southern Hemisphere extratropics.  

While the ARTMIP algorithms identify large changes in future AR frequency, the frequency of moist, poleward anomalies stays constant and relatively climate-invariant (bottom right panel of Figure~\ref{fig:djf_ar_frequency} and Appendix~Figure~A6).  A well-known future climate projection is that IWV increases at approximately 7\% per Kelvin due to the Clausius-Clapeyron relation \citep{Held2006}. Under a 7\% per Kelvin scaling, specific humidity ($q$) increases exponentially, leading to a greater increase in extreme values compared to the mean. However, when anomalies are defined relative to each climate’s own climatology, the frequency of moist anomalies ($q'^* > 0$) would be the same in both the original and scaled moisture distribution. Despite the exponential increase in moisture, the relative frequency of positive anomalies would not change. In the MRI climate simulation, the frequency of co-occurring moist, poleward anomalies does not appreciably change between historical and future climate.  The  frequency change in moist, poleward anomalies is an order of magnitude smaller than the frequency changes in the ARTMIP algorithms (Figure~\ref{fig:djf_ar_frequency}).


In the historical climate, the frequency of moist, poleward anomalies has a global mean of $\sim$28\%.  By design, this quantity is permissive: it counts any anomaly in $v'^*$ and $q'^*$ above 0, no matter how large.  Compared to the ARTMIP algorithms, the frequency of moist, poleward anomalies is more globally uniform (bottom left panel of Figure~\ref{fig:djf_ar_frequency}). It has no shape requirements or thresholds for high vapor transport.  Since the ARTMIP algorithms do have these heuristics, they tend to be more concentrated in the extratropical storm track over the oceans, and their AR frequency maps have more regional variability.

\begin{figure}[H]
\centerline{\includegraphics[width=39pc]{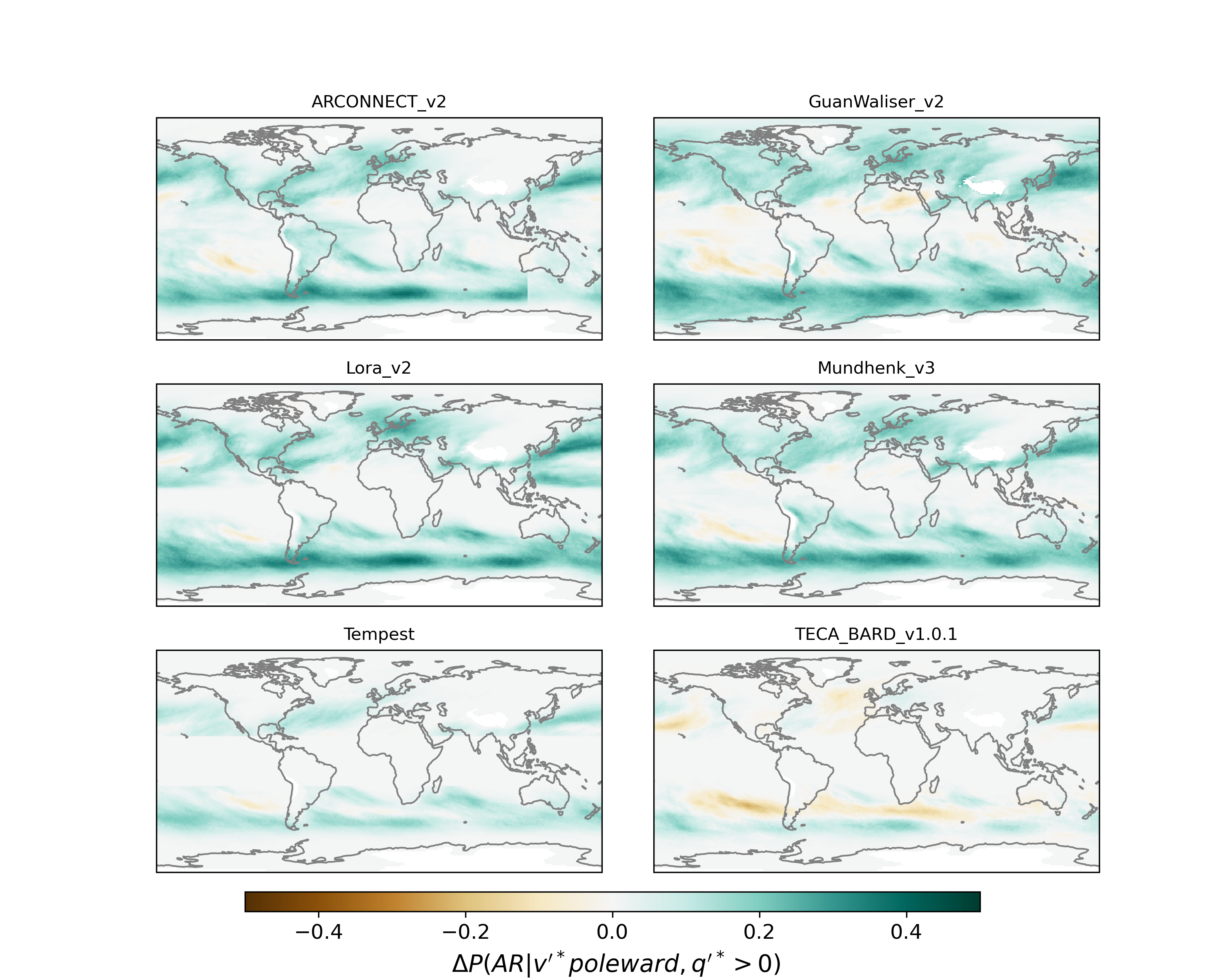}}
 \caption{\textbf{DJF Climate Changes in Conditional Probability of Atmospheric Rivers, Given Moist, Poleward Anomalies}. For six global ARTMIP detection algorithms, the change in conditional probability of an AR, given a moist, poleward $(v'^*>0 \text{ \& }q'^*>0)$ anomaly is shown.  The change is calculated for DJF between the 2081-2100 in the SSP5-8.5 and the 1995-2014 historical period simulated by the MRI-ESM2-0 climate model.}
 \label{fig:djf_conditional_ar_probability}
\end{figure}

Under climate change, a greater proportion of moist, poleward anomalies are classified as ARs (Figure~\ref{fig:djf_conditional_ar_probability} and Appendix~Figure~A7). In both DJF and JJA, the conditional probability of an AR, given a moist, poleward anomaly, increases significantly in the future climate, particularly for the GuanWaliser\_v2, Lora\_v2, and Mundhenk algorithms. However, there is still spatial variability, such as a detected decrease in AR frequency over the Pacific Ocean in DJF (though not JJA) in the ARCONNECT\_v2 algorithm.  Under future climate change, a greater proportion of moist, poleward anomalies reach the threshold to be classified as ARs. 

\section{Discussion and Conclusions}
The choice of AR detection algorithm leads to significant differences in the estimates of AR-induced LHT.  Using the \citet{Cox2024} framework to estimate instantaneous LHT, we calculate the AR contribution to TEs, SEs, and MTs.  We show the near-equal contribution of moist, poleward anomalies and dry, equatorward anomalies to TE LHT (Figure~\ref{fig:djf_lht_bound} and Appendix Figure~A2).  We show that moist, poleward anomalies are a physics-based upper bound for AR-induced LHT across all of ARTMIP. Based on the different signs of moist, poleward and dry, equatorward MTs, we physically justify the oft-cited result that ARs are responsible for the majority of extratropical moisture flux.  

We also use the upper bound as a benchmark to calculate future changes in AR intensity, variability, and frequency.  In the future under SSP5-8.5, many AR detection algorithms identify an increase in AR frequency. However, the frequency of moist, poleward anomalies remains almost constant (bottom rows of Figure~\ref{fig:djf_ar_frequency} and Appendix~Figure~A6).  Therefore, the ARTMIP algorithms classify a greater proportion of moist, poleward anomalies as ARs.  While the ARTMIP algorithms have a large spread in future AR-induced LHT, the reference AR LHT provides a clear, physics-based upper bound for the amount of LHT that can be attributed to ARs.  This is particularly useful for quantifying future changes in AR activity.  


Detection algorithms that use absolute thresholds yield different future projections than those with relative thresholds \citep{OBrien2022, Reid2025}.  In the AR reference LHT, moist, poleward anomalies are defined relative to the climatology of their respective time periods: historical anomalies are calculated with respect to the historical climatology, and future anomalies with respect to the future climatology.  This definition is rooted in atmospheric dynamics and the definition of transient eddies in the LHT decomposition.  For estimates of future AR intensity, this definition has a negligible effect on the estimated change in the magnitude of AR LHT.  The MMC accounts for a small portion of midlatitude LHT \citep{Trenberth1994}, and the change in MMC LHT is small under increased $\text{CO}_2$ \citep{Donohoe2020}.  For estimates of future AR frequency, this definition resembles a relative threshold, since it changes based on the climate scenario.  For future climate, if moist, poleward anomalies were defined with respect to historical climatology, then AR frequency would likely increase due to the moistening of the atmosphere. In future research, the climatologies and anomalies in $v$ and $q$ can also be modified to identify the contributions from thermodynamics and dynamics separately \citep{Gonzales2020, Zhou2022, Gao2015, Zavadoff2020}

In this manuscript, we only assess the relationship between ARs and meridional (north-south) LHT because this quantity is rigorously constrained by energy balance in the climate system and is well-defined from dynamic and energetic perspectives \citep{Armour2019}. However, ARs are traditionally characterized by large amounts of IVT, which also includes zonal (east-west) moisture flux.  Including zonal moisture flux estimates could yield a more comprehensive understanding of the seasonality of ARs and the total magnitude of horizontal vapor transport.

In this manuscript, we focus on the relationship between ARs and LHT because we assess the role of ARs in the global water cycle.  However, ARs have also been associated with sensible heat transport \citep{Shields2019} and surface heat extremes \citep{Scholz2024,Raymond2024}.  Future research is necessary to assess the role of ARs and moist, poleward anomalies in dry static energy transport.  This could improve understanding of the role of ARs in column-integrated moist static energy tendencies, high heat-humidity events in the midlatitudes, and polar climate change. In particular, moist and dry heat transport changes have distinct seasonality, with important implications for Arctic warming and sea ice loss \citep{Hahn2023}.  Additionally, our manuscript quantifies the relative contributions of moist, poleward anomalies; dry, equatorward anomalies; and different-signed anomalies in LHT.  A future direction of research is to quantify the analogous contributions from warm, poleward anomalies and cold, equatorward anomalies in TE dry static energy transport. This would enable a deeper understanding of the contrast between poleward and equatorward transport in a variety of climate states \citep{Caballero2012, Smith2021}.

An area of future research is to study what sets the variability in AR LHT.  In the time-mean, poleward LHT must balance incoming and outgoing solar radiation, but with substantial sub-seasonal variability, this balance is constantly perturbed. \citet{Schneider2015} show that the variability of potential temperature is proportional to the meridional gradient.  Using a Taylor expansion, they show that potential temperature anomalies are proportional to meridional displacement.  Further research is necessary to quantify this relationship for vertically integrated moisture, instead of potential temperature.  \citet{Newman2012} find that moisture variability is collocated with locations with a large meridional moisture gradient.  We find that the variability in the AR Reference LHT increases in the future (Figure~\ref{fig:ar_variability_historical_future}), and future work is necessary to assess this result in the context of a diffusive perspective of heat transport \citep{Armour2019} and a steeper meridional moisture gradient. Since the equator-to-pole gradient of moisture steepens under climate change, future work is necessary to characterize the relationship between the increase in AR reference LHT variability and the change in the meridional gradient.

Another area of future research concerns the predictability of the different eddy terms and their AR-related components.  SEs and MTs include slow-varying terms (climatological average quantities $\bar{v}^*$ and $\bar{q}^*$).  These climatological quantities may be more predictable than the instantaneous deviations, $v'^*$ and $q'^*$.  Future research is necessary to quantify the predictability of these different components and their agreement across multiple climate models.  Here, we only consider one CMIP6 climate model (the MRI climate model), since these calculations require archival of 6-hourly resolution and fully vertically resolved output.  As more high-resolution climate model simulations become available, the influence of internal variability and model uncertainty can also be quantified.

AR temporal fluctuations can have important societal consequences; they can modulate regional precipitation and runoff extremes \citep{Zhou2025} and influence the timing and intensity of climate phenomena such as column-integrated moist static energy tendencies \citep{Cox2024}. Understanding this variability offers a critical opportunity to improve both weather and climate prediction systems by explicitly accounting for the role of ARs in the global energy and moisture budgets.


%

%

\clearpage
\acknowledgments
ARTMIP is a grassroots community effort that includes a collection of international researchers from universities, laboratories, and agencies. Co-chairs and committee members include Jonathan Rutz, Christine Shields, L. Ruby Leung, F. Martin Ralph, Michael Wehner, Ashley Payne, Travis O'Brien, and Allison Collow. Details on the catalog developers can be found on the ARTMIP website (http://www.cgd.ucar.edu/projects/artmip, last access: 12 April 2025). ARTMIP has received support from the U.S. Department of Energy Office of Science Biological and Environmental Research (BER), as part of the Regional and Global Climate Modeling program, and the Center for Western Weather and Water Extremes (CW3E) at Scripps Institute for Oceanography at the University of California San Diego.

This research has been supported by the director, Office of Science, Office of Biological and Environmental Research of the U.S. Department of Energy, as part of the Regional and Global Model Analysis program area, within the Earth and Environmental Systems Modeling Program (contract no. DE-AC02-05CH11231), and used resources of the National Energy Research Scientific Computing Center (NERSC), also supported by the Office of Science of the U.S. Department of Energy (contract no. DE-AC02-05CH11231).  This research was also supported in part by the Environmental Resilience Institute, funded by Indiana University's Prepared for Environmental Change Grand Challenge initiative, and in part by Lilly Endowment, Inc., through its support for the Indiana University Pervasive Technology Institute.

%
%








%


\newpage 

 \appendix[A]

 We use the following AR detection algorithms: ar-connect \citep{Shearer2020}, guan-waliser \citep{Guan2015}, lora\_v2 \citep{Skinner2020, Lora2017}, mattingly\_v2 \citep{Mattingly2018}, mundhenk\_v3 \citep{Mundhenk2016}, panlu \citep{Pan2019}, reid \citep{Reid2020}, rutz \citep{Rutz2014}, connect \citep{Sellars2015}, sail\_v1 \citep{Gavrikov2020}, SCAFET \citep{Nellikkattil2023}, teca\_bard\_v1.0.1 \citep{OBrien2020Detection}, and tempest \citep{McClenny2020, Ullrich2021}. 

 \begin{figure}[H]
\centerline{\includegraphics[width=39pc]{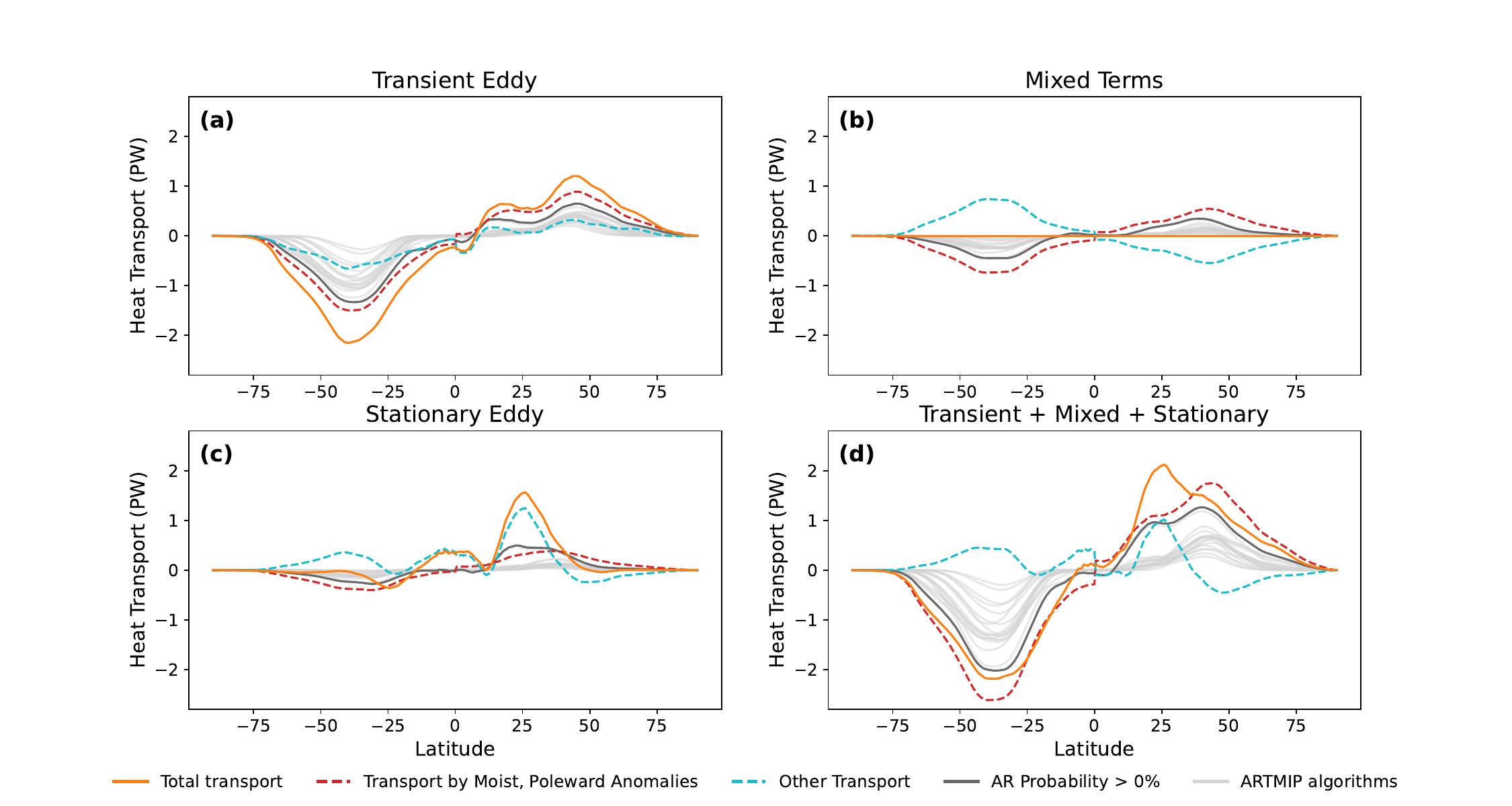}}
 \caption{\textbf{Decomposing Eddy Latent Heat Transport}. Same as Figure~\ref{fig:djf_lht_bound}, but for JJA.}
 \label{fig:jja_lht_bound}
\end{figure}

\begin{figure}[H]
\centerline{\includegraphics[width=39pc]{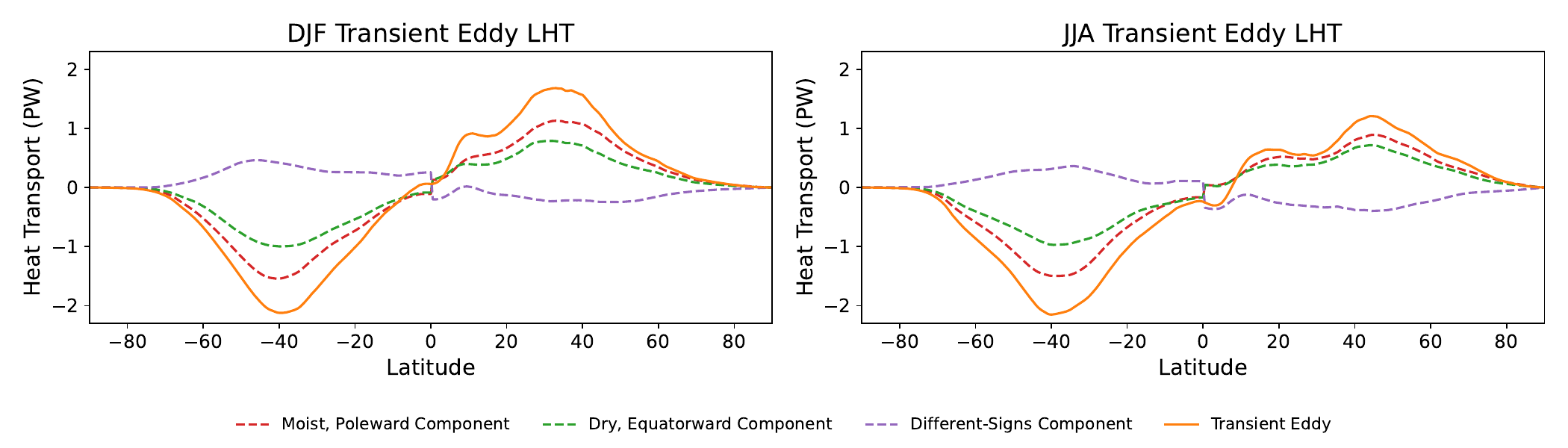}}
 \caption{\textbf{Absolute Contributions of Moist Poleward, Dry Equatorward, and Different Sign $v$ and $q$ anomalies to transient eddy transports.}  In MERRA2 (1995-2014), the contributions of moist, poleward; dry, equatorward; and different-sign anomalies are shown to transient eddy LHT.  Contributions are shown for DJF (left) and JJA (right).  See Figure~\ref{fig:relative_contributions_mp_de} for the relative contributions.}
 \label{fig:absolute_contributions_mp_de}
\end{figure}

\begin{figure}[H]
\centerline{\includegraphics[width=39pc]{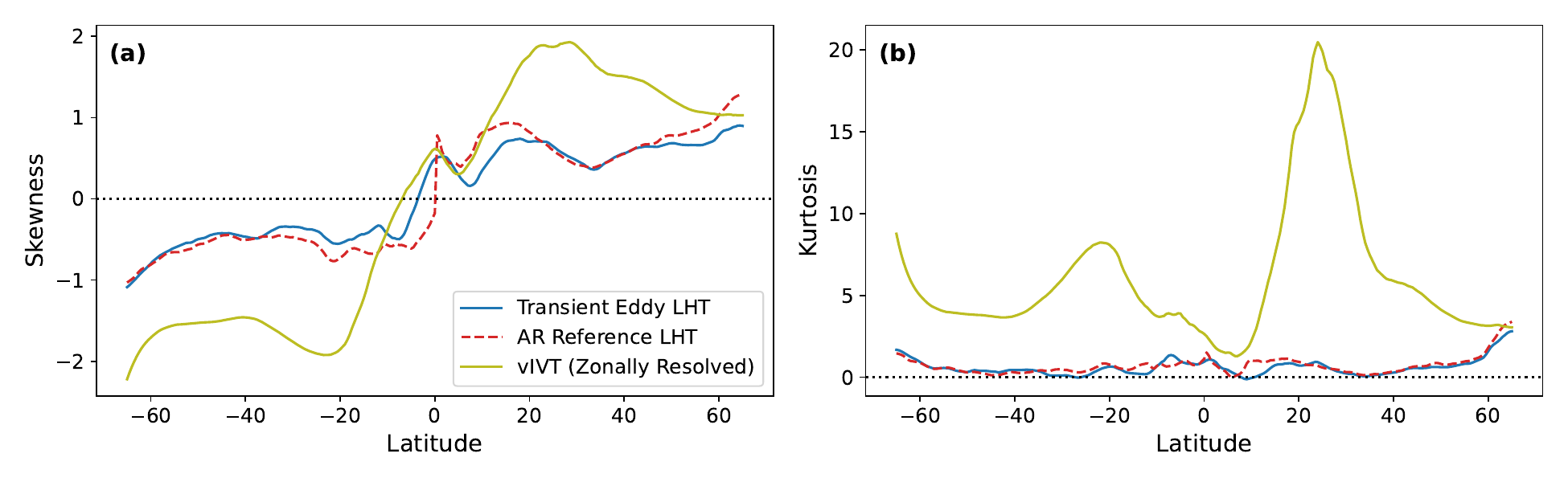}}
 \caption{\textbf{JJA Temporal Distribution of total LHT, transient eddy LHT, and AR Upper Bound LHT.}  Same as Figure~\ref{fig:djf_skewness_kurtosis} but for JJA.}
 \label{fig:jja_skewness_kurtosis}
\end{figure}

\begin{figure}[H]
\centerline{\includegraphics[width=39pc]{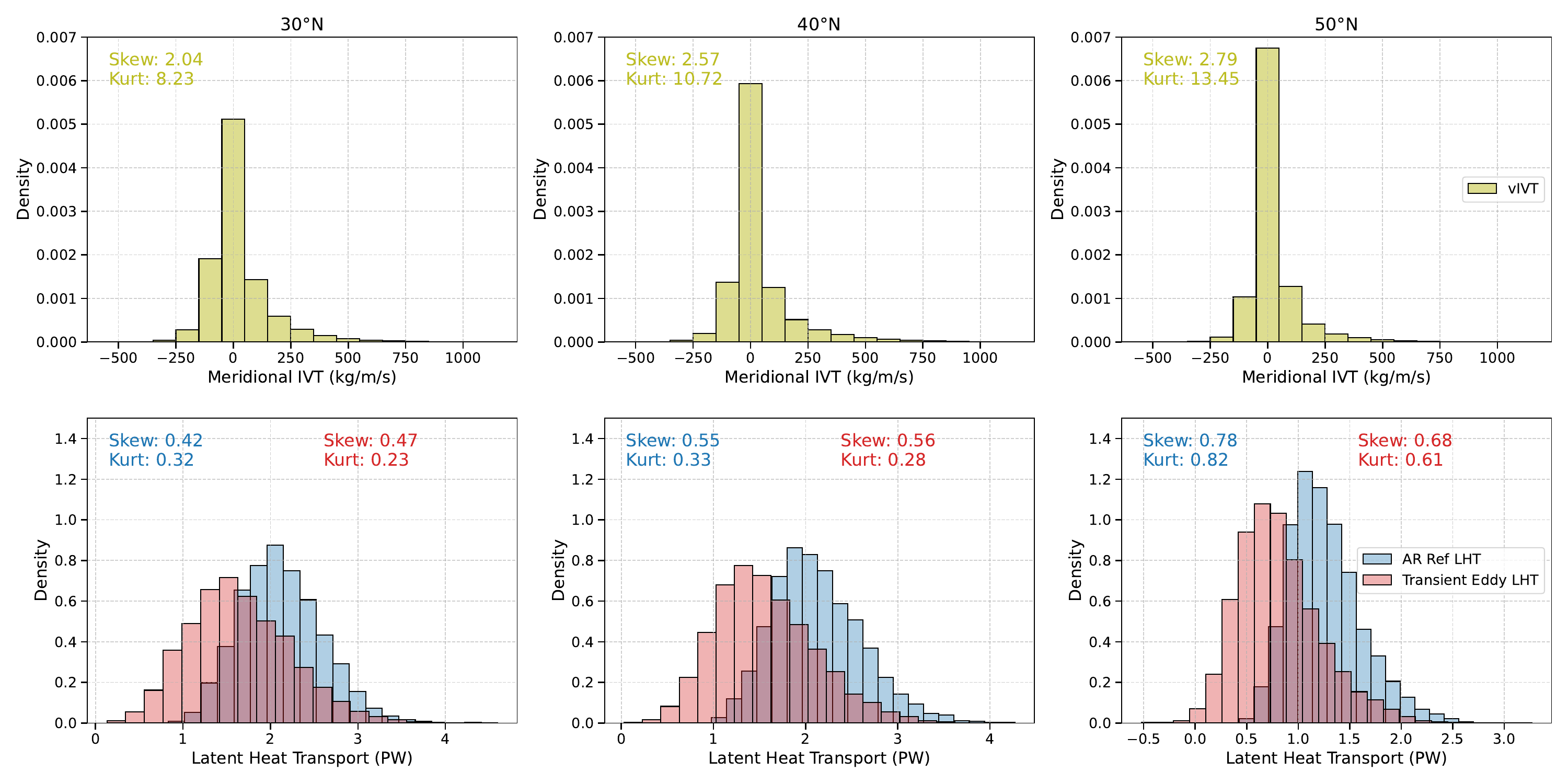}}
 \caption{\textbf{Histograms of vIVT (top row) and AR reference LHT and TE LHT (bottom row)}.  Visualization of distributions corresponding to the skewness and kurtosis values shown in Figure~\ref{fig:djf_skewness_kurtosis}).  Top row shows vIVT, which is a zonally resolved quantity.  Bottom row shows AR reference LHT and TE LHT, which are zonal mean quantities.  Skewness and kurtosis values for each distribution are printed alongside them.}
 \label{fig:distribution_vis}
\end{figure}

\begin{figure}[H]
\centerline{\includegraphics[width=39pc]{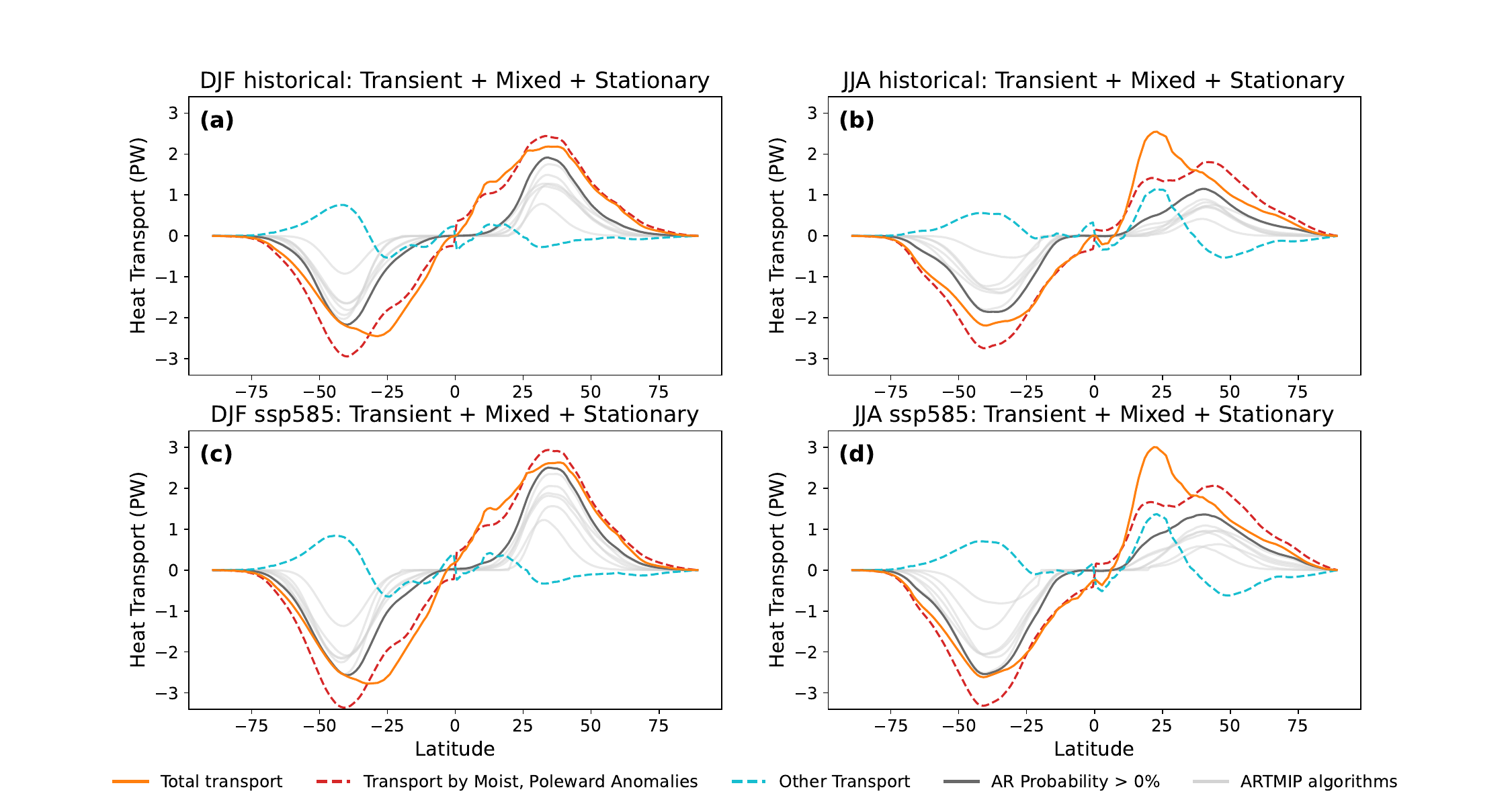}}
 \caption{\textbf{Decomposing Eddy Latent Heat Transport in MRI-ESM2-0.}  The sum of transient eddies, stationary eddies, and mixed terms are shown for DJF and JJA in MRI-ESM2-0 historical (1995-2014) and SSP5-8.5 (2081-2100).}
 \label{fig:mri_lht_bound}
\end{figure}

\begin{figure}[H]
\centerline{\includegraphics[width=21pc]{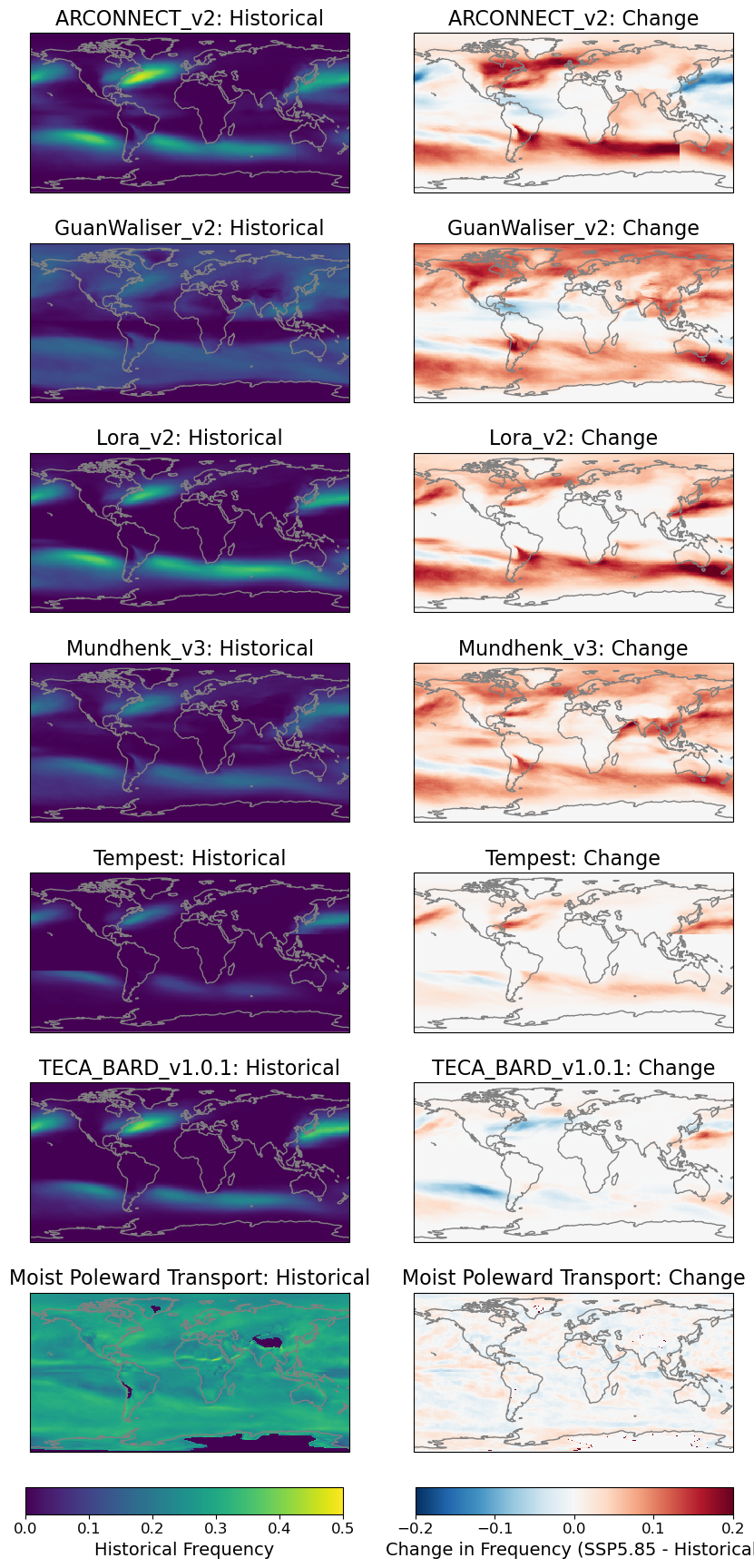}}
\caption{\textbf{Climate Changes in Frequency of ARs and Moist, Poleward Anomalies.} Same as Figure~\ref{fig:jja_ar_frequency} but for JJA.}
 \label{fig:jja_ar_frequency}
\end{figure}

\begin{figure}[H]
\centerline{\includegraphics[width=39pc]{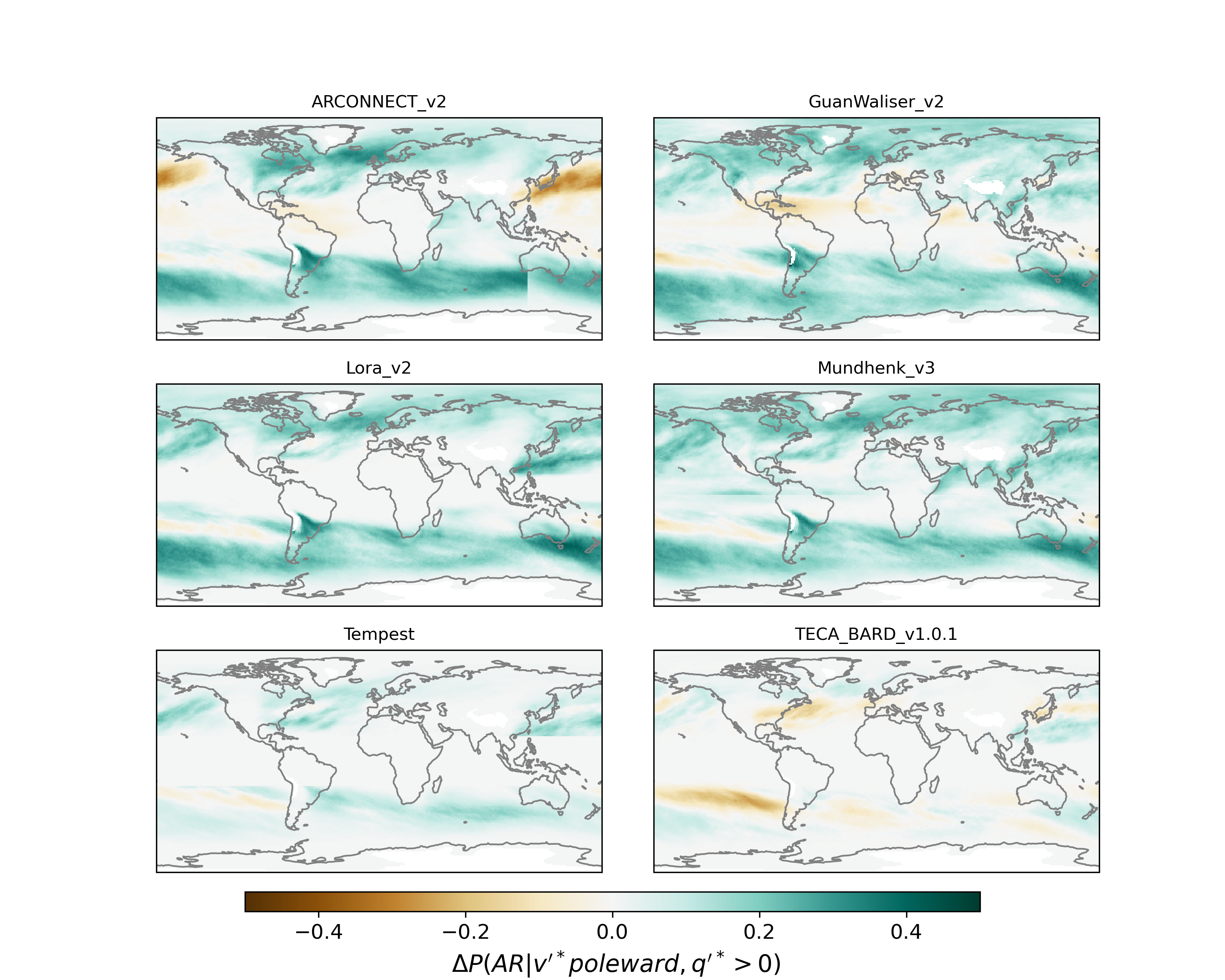}}
 \caption{\textbf{Climate Changes in Conditional Probability of Atmospheric Rivers, Given Moist, Poleward
Anomalies.}  Same as Figure~\ref{fig:djf_conditional_ar_probability} but for JJA.}
 \label{fig:jja_conditional_ar_probability}
\end{figure}

\newpage
\appendix[B]



\section{An algorithm to calculate MT LHT from moist, poleward anomalies}

Algorithm~3 lists the method to calculate the contribution of moist, poleward anomalies to MT LHT.

\begin{algorithm}[H]
\caption{Computation of Moist, Poleward Mixed Term Latent Heat Transport (MT LHT)}
\label{alg:ar_mt_lht}
\begin{algorithmic}[1]

\Statex \textbf{Input:} \begin{itemize}[noitemsep, topsep=0pt]
    \item $v(\theta,\phi,p,t)$
    \item $q(\theta,\phi,p,t)$
    \item Heaviside mask, $H(\tilde{v}'^*)$, for poleward anomalies
    \item Heaviside mask, $H(q'^*)$, for moist anomalies
\end{itemize} 
\Statex \textbf{Output:}  Contribution of Moist, Poleward Anomalies to MT LHT 

\State Compute $v'^*_\text{MP}$ and $q'^*_\text{MP}$ using Equations~\ref{eq:vprime_decomp} and \ref{eq:qprime_decomp} (Algorithm~\ref{alg:ar_te_lht}), respectively.

\State Compute $\overline{v}^*_\text{MP}$ and $\overline{q}^*_\text{MP}$ using  Equations~\ref{eq:vprimebar_decomp} and \ref{eq:qprimebar_decomp} (in Algorithm~\ref{alg:ar_se_lht}), respectively.  The time-averaging is performed after the Heaviside function.

\State Calculate the MT LHT:
\begin{align}
\text{Moist, Poleward MT LHT}(\theta) &= \frac{2\pi a \cos(\theta)}{g} L_v \int_0^{P_s} \big ([\bar{v}^*_{\text{MP}} q'^*_{\text{MP}}] + \notag \\  &
\quad\quad\quad\quad\quad\quad\quad[v'^*_{\text{MP}} \bar{q}^*_{\text{MP}}] \big ) \,dp \label{eq:mp_se_lht}
\end{align}

\end{algorithmic}
\end{algorithm}

\section{Decomposition of TE LHT into 3 sub-terms}

First, by definition, 

\begin{align*}
1 - H(\tilde{v}'^*) = H(-\tilde{v}'^*) \\
1 - H(q'^*) = H(-q'^*)
\end{align*}

Therefore,

\begin{align}
1 - H(\tilde{v}'^*)H(q'^*)  &=  (1-H(\tilde{v}'^*))H(q'^*) + H(\tilde{v}'^*)(1 - H(q'^*)) + (1 - H(\tilde{v}'^*)) (1 - H(q'^*)) \notag \\ &= H(-\tilde{v}'^*) H(q'^*) + H(\tilde{v}'^*)H(-q'^*) + H(-\tilde{v}'^*)H(-q'^*) \label{eq:heaviside_expansion_appdx}
\end{align}


Based on Equations~\ref{eq:vprime_decomp}, ~\ref{eq:qprime_decomp}, and Equation~\ref{eq:heaviside_expansion_appdx}

\begin{align}
    v'^* & = v'^*H(\tilde{v}'^*)H(q'^*) + v'^*(1-H(\tilde{v}'^*)H(q'^* )) \notag \\ &
    = \underbrace{v'^*H(\tilde{v}'^*)H(q'^*)}_{\text{Moist, Poleward}} + \underbrace{v'^*H(-\tilde{v}'^*)H(-q'^*)}_{\text{Dry, Equatorward}} + \underbrace{v'^*H(-\tilde{v}'^*)H(q'^*) + v'^*H(\tilde{v}'^*)H(-q'^*)}_{\text{Different-Signs}} \notag\\  &
     = v'^*_{\text{MP}} + v'^*_{\text{DE}} + \underbrace{v'^*_\text{ME} + v'^*_\text{DP}}_{\text{Different-Signs}} 
\end{align}

Similarly for q: 
\begin{align}
    q'^* & = q'^*H(\tilde{v}'^*)H(q'^*) + q'^*(1-H(\tilde{v}'^*)H(q'^* )) \notag \\ &
    = \underbrace{v'^*H(\tilde{v}'^*)H(q'^*)}_{\text{Moist, Poleward}} + \underbrace{q'^*H(-\tilde{v}'^*)H(-q'^*)}_{\text{Dry, Equatorward}} + \underbrace{q'^*H(-\tilde{v}'^*)H(q'^*) + v'^*H(\tilde{v}'^*)H(-q'^*)}_{\text{Different-Signs}} \notag\\  &
     = q'^*_{\text{MP}} + q'^*_{\text{DE}} + \underbrace{q'^*_\text{ME}+q'^*_\text{DP}}_{\text{Different-Signs}} 
\end{align}

Instantaneous TE LHT can be written as 

\begin{align}
\text{Instantaneous TE LHT}(\theta) &= \frac{2\pi a \cos(\theta)}{g} L_v \int_0^{P_s} [v'^* q'^*] \, dp \\ 
&= \frac{2\pi a \cos(\theta)}{g} L_v \int_0^{P_s} [(v'^*_\text{MP} + v'^*_\text{DE} + \underbrace{v'^*_\text{ME } + v'^*_\text{DP }}_{\text{Different-Signs}} \big ) \notag \\ &
\quad\quad\quad\quad\quad\quad(q'^*_\text{MP} + q'^*_\text{DE} + \underbrace{q'^*_\text{ME} + q'^*_\text{DP}}_{\text{Different-Signs}}  \big )] \, dp \label{eq:te_decomp_appendix}
\end{align}

$H(\tilde{v}'^*)H(-\tilde{v}'^*)$ is identically zero.  $H(q'^*)H(-q'^*)$ is identically zero.  Therefore, cross terms in Equation~\ref{eq:te_decomp_appendix}, such as $v'^*_{\text{MP}}q'^*_{\text{DE}} = v'^*H(\tilde{v}'^*)H(q'^*)q'^*H(-\tilde{v}'^*)H(-q'^*)$ equal 0.  A similar argument applies for all cross terms: $v'^*_{\text{MP}} q'^*_{\text{DE}}$, $v'^*_{\text{MP}} q'^*_{\text{ME}}$, $v'^*_{\text{MP}} q'^*_{\text{DP}}$, $v'^*_{\text{DE}} q'^*_{\text{MP}}$, 
$v'^*_{\text{DE}} q'^*_{\text{ME}}$,
$v'^*_{\text{DE}} q'^*_{\text{DP}}$,
$v'^*_{\text{ME}} q'^*_{\text{MP}}$,
$v'^*_{\text{DP}} q'^*_{\text{MP}}$, and $v'^*_{\text{ME}} q'^*_{\text{DE}}$,
$v'^*_{\text{DP}} q'^*_{\text{DE}}$.  Therefore, Equation~\ref{eq:te_decomp_appendix} reduces to three terms:

\begin{align}
\text{Instantaneous TE LHT}(\theta) 
&= \frac{2 \pi a \cos \theta}{g} L_v \int_0^{P_s} [v'^*q'^*] dp \\
&= \frac{2 \pi a \cos \theta}{g} L_v \int_0^{P_s}\bigg( 
    \underbrace{[v'^*_{\text{MP}}q'^*_{\text{MP}}]}_{\text{Moist, Poleward}} 
    + \nonumber \\ &\quad\quad\quad\quad \underbrace{[v'^*_{\text{DE}}q'^*_{\text{DE}}]}_{\text{Dry, Equatorward}} + \nonumber \\
&\quad\quad\quad\quad
    \underbrace{[v'^*_{\text{ME}}q'^*_{\text{ME}}] + [v'^*_{\text{DP}}q'^*_{\text{DP}}]}_{\text{Different-Signs}}
\bigg) dp
\end{align}

These terms are the "moist, poleward" component, "dry, equatorward" component, and "different-signs" component.  In Figure~\ref{fig:djf_lht_bound}a, the moist, poleward component is shown in red, and the sum of the dry, equatorward component and the other component is shown in blue.  Since instantaneous TE LHT reduces to three terms, the relative magnitudes of the moist, poleward; dry, equatorward; and other components are shown in Figure~\ref{fig:relative_contributions_mp_de}.

\section{Decomposition of SE and MT LHT into 9 sub-terms}

Based on Equations~\ref{eq:vprimebar_decomp}, ~\ref{eq:qprimebar_decomp}, and \ref{eq:heaviside_expansion_appdx},

\begin{align}
    \bar{v}^* & = \overline{v^* H(\tilde{v}'^*) H(q'^*)} + \overline{v^* \big(1 - H(\tilde{v}'^*) H(q'^*) \big)} \notag \\ 
    & = \underbrace{\overline{v^* H(\tilde{v}'^*) H(q'^*)}}_{\text{Moist, Poleward}} 
       + \underbrace{\overline{v^* H(-\tilde{v}'^*) H(-q'^*)}}_{\text{Dry, Equatorward}} 
       + \underbrace{\overline{v^* H(-\tilde{v}'^*) H(q'^*) + v^* H(\tilde{v}'^*) H(-q'^*)}}_{\text{Different-Signs}} \notag \\  
    & = \bar{v}^*_{\text{MP}} + \bar{v}^*_{\text{DE}} + \underbrace{\bar{v}^*_\text{ME}+\bar{v}^*_\text{DP}}_{\text{Different-Signs}}
\end{align}
Similarly for q:
\begin{align}
    \bar{q}^* & = \overline{q^* H(\tilde{v}'^*) H(q'^*)} + \overline{q^* \big(1 - H(\tilde{v}'^*) H(q'^*) \big)} \notag \\ 
    & = \underbrace{\overline{q^* H(\tilde{v}'^*) H(q'^*)}}_{\text{Moist, Poleward}} 
       + \underbrace{\overline{q^* H(-\tilde{v}'^*) H(-q'^*)}}_{\text{Dry, Equatorward}} 
       + \underbrace{\overline{q^* H(-\tilde{v}'^*) H(q'^*) + q^* H(\tilde{v}'^*) H(-q'^*)}}_{\text{Different-Signs}} \notag \\  
    & = \bar{q}^*_{\text{MP}} + \bar{q}^*_{\text{DE}} + \underbrace{\bar{q}^*_\text{ME}+\bar{q}^*_\text{DP}}_{\text{Different-Signs}}
\end{align}

The SE LHT is calculated as

\begin{align}
\text{SE LHT}(\theta) &= \frac{2\pi a \cos(\theta)}{g} L_v \int_0^{P_s} [\bar{v}^* \bar{q}^*] \, dp \\ 
&= \frac{2\pi a \cos(\theta)}{g} L_v \int_0^{P_s} [(\bar{v}^*_\text{MP} + \bar{v}^*_\text{DE} + \underbrace{\bar{v}^*_\text{ME} + \bar{v}^*_\text{DP}}_{\text{Different-Signs}} \big ) \notag (\bar{q}^*_\text{MP} + \\ &
\quad\quad\quad\quad\quad\quad\quad\quad \bar{q}^*_\text{DE} + \underbrace{\bar{q}^*_\text{ME} + \bar{q}^*_\text{DP}}_{\text{Different-Signs}} \big )] \, dp \label{eq:se_decomp_appendix}
\end{align}

For the time-mean terms $\bar{v}^*$ and $\bar{q}^*$, the Heaviside function is applied before the time averaging.  Therefore, the resulting cross terms (e.g. $\bar{v}^*_\text{MP}\bar{q}^*_\text{DE}$) do not necessarily equal 0.

\begin{align}
    \bar{v}^*_\text{MP}\bar{q}^*_\text{DE} = \overline{v^*H(\tilde{v}'^*)H(q'^*))} \, \overline{q^*H(-\tilde{v}'^*)H(-q'^*)} \neq 0
\end{align} 

This means that all 9 terms from Equation~\ref{eq:se_decomp_appendix} are retained in the SE decomposition.

\newpage
\bibliographystyle{ametsocV6}
\bibliography{references}

\end{document}